\documentclass[useAMS,usenatbib]{mn2e}
\usepackage{graphicx}
\usepackage{amsmath}
\usepackage{amsfonts}
\usepackage{amssymb}
\usepackage{rotating}
\usepackage[english]{babel}

\title{Detonations in white dwarf dynamical interactions}

\author[G. Aznar-Sigu\'an et al.]{G. Aznar-Sigu\'an$^{1,2}$,
                                 E. Garc\'\i a--Berro$^{1,2}$,
                                 P. Lor\'en--Aguilar$^{3}$,
                                 J. Jos\'e$^{4,2}$ and 
                                 J. Isern$^{5,2}$\\
       $^1$Departament de F\'\i sica Aplicada, 
           Universitat Polit\`ecnica de Catalunya,
           c/Esteve Terrades 5, 
           08860 Castelldefels, Spain\\
       $^2$Institute for Space  Studies of Catalonia,
           c/Gran Capit\`a 2--4, Edif. Nexus 104,   
           08034  Barcelona, Spain\\
       $^3$School of Physics, University of Exeter, 
           Stocker Road, Exeter, 
           UK EX4 4QL, United Kingdom\\
       $^4$Departament de F\'\i sica i Enginyeria Nuclear,
           Universitat Polit\`ecnica de Catalunya,
           c/Comte d'Urgell 187,
           08036 Barcelona, Spain\\
       $^5$Institut de Ci\`encies de l'Espai, CSIC,  
           Campus UAB, Facultat de Ci\`encies, Torre C-5, 
           08193 Bellaterra, Spain}

\begin{document}

\date{\today}

\maketitle

\begin{abstract}
In old, dense stellar systems  collisions of white dwarfs are a rather
frequent phenomenon.   Here we present the results  of a comprehensive
set of Smoothed Particle Hydrodynamics simulations of close encounters
of white  dwarfs aimed to explore  the outcome of  the interaction and
the  nature of the  final remnants  for different  initial conditions.
Depending on the initial conditions  and the white dwarf masses, three
different  outcomes are  possible.  Specifically,  the outcome  of the
interaction  can be  either a  direct or  a lateral  collision  or the
interaction can result in the formation of an eccentric binary system.
In those cases in which  a collision occurs, the infalling material is
compressed  and  heated  such  that  the  physical  conditions  for  a
detonation  may be  reached  during  the most  violent  phases of  the
merger.  While we find that  detonations occur in a significant number
of our  simulations, in some of  them the temperature  increase in the
shocked region  rapidly lifts degeneracy, leading to  the quenching of
the  burning.   We  thus  characterize  under  which  circumstances  a
detonation  is likely  to  occur as  a  result of  the  impact of  the
disrupted  star  on the  surface  of  the  more massive  white  dwarf.
Finally, we also study which interactions result in bound systems, and
in which  ones the  more massive  white dwarf is  also disrupted  as a
consequence  of  the dynamical  interaction.   The  sizable number  of
simulations performed in  this work allows to find  how the outcome of
the interaction  depends on the  distance at closest approach,  and on
the masses  of the colliding white  dwarfs, and which  is the chemical
pattern of the nuclearly processed material.  Finally, we also discuss
the  influence of  the masses  and core  chemical compositions  of the
interacting  white dwarfs  and the  different kinds  of impact  in the
properties of the remnants.
\end{abstract}



\section{Introduction}

In globular clusters, the density  of stars is roughly a million times
that of  our Solar System's  environs.  In such dense  stellar systems
stellar collisions  are rather frequent  \citep{HiDa76}.  Actually, it
has been predicted that up to 10\% of the stars in the core of typical
globular clusters have undergone a  collision at some point during the
lifetime  of the  cluster  \citep{Da02}. Also,  galactic nuclei  which
harbor massive black holes, like  that of our own Galaxy, have stellar
densities  at least  as large  as  those found  in the  center of  the
densest  globular  clusters.   Moreover,  in  these  environments  the
frequency  of stellar collisions  is strongly  enhanced, since  due to
strong attraction of their central black holes, stars have much larger
velocities.

In these very  dense stellar systems the most  probable collisions are
those  in which  at  least one  of  the colliding  stars  has a  large
cross-section --- a red giant or an AGB star --- and those in which at
least one  of the stars  is common \citep{ShRe86}. Since  white dwarfs
are the  most usual end-point  of stellar evolution, and  because both
globular clusters  and galactic nuclei  are rather old,  these stellar
systems contain many degenerate stars.  Therefore, collisions in which
one of  the colliding stars is  a white dwarf should  be rather common
\citep{PaEn10}.  Additionally,  it has been recently  shown that close
encounters of two white dwarfs  could be more frequent than previously
thought \citep{KD12}.

Recently, the study of the collisions of two white dwarfs has received
considerable  interest, since it  has been  shown that,  under certain
circumstances,  the result  of such  interactions could  be a  Type Ia
supernova  outburst  \citep{RaSc09,RaSc10,RoKa09}.   Moreover, it  has
also been suggested that such processes could not only lead to Type Ia
supernova  explosions,   but  also  to  the   formation  of  magnetars
\citep{KiPr01}, and could also be at the origin of high-field magnetic
white  dwarfs  \citep{hfmwd}.   Additionally,  this  is  a  suggestive
scenario which could also explain  some of the characteristics of soft
gamma-ray repeaters and of  anomalous X-ray pulsars \citep{Rueda}.  It
is also  interesting to note  that dynamical interactions  in globular
clusters can  form double  white dwarfs with  non-zero eccentricities.
These   systems   are   powerful   sources  of   gravitational   waves
\citep{WiKa07},  which  could  be  eventually detected  by  spaceborne
observatories \citep{LAea05}.  Finally,  another important property of
the collision  of white  dwarfs is that  the temperatures  achieved in
direct collisions are substantially high, and consequently some of the
nuclearly  processed  material ejected  during  the interaction  could
pollute the surrounding environment \citep{PaEn10}.

All in all, it is clear that for multiple reasons the collision of two
white  dwarfs   is  a   subject  that  deserves   to  be   studied  in
detail. However,  there are very  few simulations of  this phenomenon.
In fact, the small number  of simulations of white dwarf collisions is
noticeable when  compared to  the number of  simulations in  which two
white dwarfs  belonging to a  binary system merge.   Specifically, the
first simulations  of colliding  white dwarfs \citep{Bea89}  were done
using a Smoothed Particle  Hydrodynamics (SPH) method, but employing a
rather small number of particles in the calculations, a consequence of
the  severe  computational  limitations.   Other  recent  calculations
\citep{RoKa09}  also employed  SPH techniques,  but due  to  the large
parameter space to be studied, the calculations were mostly restricted
to head-on  collisions, this time  using a large number  of particles.
The reason for  this choice was that these  simulations were primarily
aimed at obtaining a thermonuclear explosion, and it was foreseen that
very high temperatures  were most likely to be  obtained in these kind
of  interactions.  This was  also the  aim of  subsequent calculations
\citep{RaSc09,RaSc10},  which  indepently  confirmed  the  results  of
\cite{RoKa09} using an independent  SPH code.  Finally, the collisions
of two white dwarfs have also been studied recently using the Eulerian
adaptive  grid code  FLASH \citep{Hawley12}.   However,  these authors
only  computed zero  impact  parameter collisions  for two  equal-mass
white  dwarfs, a  $0.64\,  M_{\sun} +  0.64  \, M_{\sun}$  pair and  a
$0.81\,  M_{\sun} +  0.81  \, M_{\sun}$  system.   As in  most of  the
previous studies of this kind, the primary goal of this work was again
to study an independent channel for producing Type Ia supernovae.

In  summary, most  authors have  studied  the collision  of two  white
dwarfs with fixed masses, and have varied the total energy and angular
momentum  of the colliding  white dwarfs,  while little  attention has
been  paid up  to  now  to study  the  effects of  the  masses of  the
interacting   stars.   This   was   also  the   approach  adopted   by
\cite{PaEn10},  where  the  masses  of the  intervening  white  dwarfs
($0.8\, M_{\sun}$  and $0.6\, M_{\sun}$) were kept  fixed, while their
initial relative  velocity and distance were varied.   They found that
the outcome of  the interaction could be either  a direct collision, a
lateral one, or could be  the formation of an eccentric binary system.
Nevertheless, the outcome of  the dynamical interaction depends on the
masses (and on the core chemical composition) of the interacting white
dwarfs.  Thus, a comprehensive study  of the interactions of two white
dwarfs  covering a  broad range  of masses  and initial  conditions is
still lacking.  The present work aims precisely at filling this gap.

Our  paper is  organized as  follows. In  Sect.~\ref{input}  the input
physics  and  the  method  of  calculation  used  in  the  simulations
described here are briefly explained.  We pay special attention to the
numerical technique  employed in  our calculations, the  so-called SPH
method,  and   to  our  current  implementation   of  such  technique.
Sect.~\ref{ini} is  devoted to present the initial  conditions used in
this work.  It follows  Sect.~\ref{results}, where the outcomes of the
collisions  and  close  encounters  are presented  and  analyzed.   In
Sect.~\ref{remnants} we  present the most significant  features of the
extensive  set   of  simulations   performed  so  far.    Finally,  in
Sect.~\ref{concl}  we summarize  our main  findings, we  discuss their
significance and we draw our conclusions.

\section{Input physics and method of calculation}
\label{input}

As mentioned  earlier, the  hydrodynamic evolution of  the interacting
white dwarfs has been followed using  a SPH code.  SPH is a Lagrangian
particle numerical  method. It was first proposed  by \cite{Lu77} and,
independently, by  \cite{GiMo77}.  The basic principle  of SPH methods
consists in discretizing  the fluid in a set  of elements, referred to
as particles.  These particles have  a spatial dimension (known as the
``smoothing length'',  $h$), over which their  properties are smoothed
using a  kernel.  In the following  we explain the  main features that
characterize our SPH code.

Our SPH code  is fully parallel, and allows to  run simulations with a
large number of  particles in a reasonable time.   We use the standard
cubic spline  kernel of \cite{MoLa85}.   The gravitational interaction
is  also softened  using this  kernel \citep{HeKa89}.   The  search of
neighbors and the evaluation of the gravitational forces are performed
using an octree \citep{BaHu86}.  This is a tree method which allows to
directly calculate  the force  in the gravitational  $N$-body problem.
The computational  load of  this method grows  only as $N\log  N$.  To
determine  the smoothing  length of  each particle  we use  a standard
prescription:
\begin{equation}
h_i=\eta\left(\frac{m_i}{\rho_i}\right)^{1/3}
\end{equation}
We also consider the effect of the gradient of the smoothing length on
the SPH equations, which is given by:

\begin{equation}
\Omega_i=\left(1-\frac{\partial h_i}{\partial\rho_i}\sum_j m_j 
\frac{\partial W_{ij}(h_i)}{\partial h_i}\right).
\end{equation}

To deal numerically with shocks,  where on the macroscopic scales of a
simulation   the  very  steep   gradients  appear   as  discontinuous,
artificial viscosity is  usually added to the SPH  equations.  The SPH
code  employed  in  the   simulations  uses  a  prescription  for  the
artificial viscosity based in Riemann-solvers \citep{Mo97}:
\begin{equation}
 \Pi_{ij}=-\alpha\frac{v_{ij}^{\rm sig}}{\overline{\rho}_{ij}} 
\mathbf{v}_{ij} \cdot \hat{e}_{ij}
\label{av}
\end{equation}
where  the   signal  velocity   is  taken   as  $v^{\rm   sig}_{ij}  =
\overline{c}_{ij}            -\min\{0,\mathbf{v}_{ij}            \cdot
\hat{\mathbf{e}}_{ij}\}$        and        $\overline{c}_{ij}        =
({cs}_{i}+{cs}_{j})/2$  is  the  averaged   sound  speed.   The  other
variables   have   their   usual  meaning:   $\overline{\rho}_{ij}   =
({\rho}_{i}+{\rho}_{j})/2$,              $\mathbf{v}_{ij}=\mathbf{v}_i
-\mathbf{v}_j$,        $\hat{\mathbf{e}}_{ij}=        \mathbf{r}_{ij}/
|\mathbf{r}_{ij}|$,  $\mathbf{r}_{ij}= \mathbf{r}_i-\mathbf{r}_j$  and
$\alpha$  is an  adjustable parameter.  Normally, $\alpha=0.5$  yields
good results.   Additionally, to suppress artificial  viscosity forces
in pure shear flows the viscosity switch of \cite{Ba95} is also used:
\begin{equation*}
f_i=\frac{\left|\nabla\cdot\mathbf{v}\right|_i}
{\mid\nabla\cdot\mathbf{v}\mid_i+\mid\nabla\times\mathbf{v}\mid_i+{10}^{-4} 
c_i / h_i}
\end{equation*}
In this way the dissipative terms are largely reduced in most parts of
the fluid and are only used where they are really necessary to resolve
a shock, if present.

Within this  approach, the SPH  equations for the momentum  and energy
conservation read respectively:

\begin{eqnarray}
\frac{d\mathbf{v}_i}{dt}  & = & - \sum_j m_j \Big[ \frac{P_i}
{\rho_i^2\Omega_i}F_{ij}(h_i) + \frac{P_j}{\rho_j^2\Omega_j}F_{ij}(h_j) 
\nonumber \\
 & + & \overline{f}_{ij}\frac{\Pi_{ij}}{2}\left(\frac{1}{\Omega_i}F_{ij}(h_i)
+\frac{1}{\Omega_j}F_{ij}(h_j)\right) \Big]\mathbf{r}_{ij} - \nabla\Phi_i
\label{mom}
\end{eqnarray}

\begin{eqnarray}
\frac{du_i}{dt} &=& \frac{1}{\Omega_i}\sum_j m_j F_{ij}(h_i)
\left(\frac{P_i}{\rho_i^2} + \frac{\Pi_{ij}}{2}\overline{f}_{ij} \right) 
\mathbf{v}_{ij} \cdot \mathbf{r}_{ij} + \varepsilon_i 
\label{uint}
\end{eqnarray}

\noindent  where  we have  used  a  function  $F(r/h)$ such  that  the
gradient    of     the    kernel    is     $\nabla_i    W_{ij}(h)    =
F_{ij}(h)\mathbf{r}_{ij}$.   We  also  took  into account  the  energy
released by nuclear reactions, $\varepsilon_i$.

We found that it is sometimes advisable to use a different formulation
of the equation of  energy conservation \citep{Gea04, LAea05, LAea09}.
Accordingly, for each timestep the variation of the internal energy is
computed  using Eq.~(\ref{uint}) and  simultaneously the  variation of
temperature is computed using:
\begin{eqnarray}
\frac{dT_{i}}{dt}  & = & \frac{1}{\Omega_i}\sum_{j=1}^{N} \frac{m_j}{{C_v}_i} 
\left(\frac{T_i}{\rho_i^2}
\left[\left(\frac{\partial P}{\partial T}\right)_\rho\right]_i 
 + \frac{\Pi_{ij}}{2}\overline{f}_{ij}\right)\nonumber \\
&\, & \mathbf{v}_{ij} \cdot \mathbf{r}_{ij} F_{ij}(h_i) + \frac{\varepsilon_i}
{{C_v}_i}
\label{temp}
\end{eqnarray}
where $C_v$ is the specific heat.   In order to avoid errors made when
computing  the  temperature from  the  internal  energy in  degenerate
regions,   we   compare   the   temperatures   obtained   when   using
Eqs.~(\ref{uint})  and (\ref{temp}).  If  the difference  between both
values  is larger  than $5\%$,  we consider  the  temperature obtained
using Eq.~(\ref{temp}). Otherwise, we follow the temperature evolution
of the SPH particles  using Eq.~(\ref{uint}).  Using this prescription
energy is best conserved. 

Regarding  the  integration  method, a  predictor-corrector  numerical
scheme with  variable timestep \citep{SeCh96},  which turns out  to be
quite accurate, is used. The sequence initiates by predicting variable
values (denoted by primes) at $t_{n+1}$ according to

\begin{eqnarray*}
 \mathbf{r}'_{n+1} &=& \mathbf{r}_n + \mathbf{v}_n \Delta t_n + 
 \mathbf{a}'_n (\Delta t_n)^2/2 \\
 \mathbf{v}'_{n+1} &=& \mathbf{v}_n + \mathbf{a}'_n \Delta t_n \\
 \mathbf{u}'_{n+1} &=& \mathbf{u}_n + \mathbf{\dot{u}}'_n \Delta t_n \\ 
 \mathbf{T}'_{n+1} &=& \mathbf{T}_n + \mathbf{\dot{T}}'_n \Delta t_n 
\end{eqnarray*}

The   above   predicted   quantities   are  then   used   to   compute
$\mathbf{a}'_{n+1}$,           $\mathbf{\dot{u}}'_{n+1}$           and
$\mathbf{\dot{T}}'_{n+1}$       at      $\mathbf{r_{n+1}}'$      using
Eqs.~(\ref{mom}),   (\ref{uint})  and  (\ref{temp}).    The  predicted
quantities are then used to correct the positions, velocities, thermal
energies and temperatures:

\begin{eqnarray*}
 \mathbf{r}_{n+1} &=& \mathbf{r}'_{n+1} + A(\mathbf{a}'_{n+1} - 
 \mathbf{a}'_{n})(\Delta t_n)^2/2 \\
 \mathbf{v}_{n+1} &=& \mathbf{v}'_{n+1} + B(\mathbf{a}'_{n+1} - 
 \mathbf{a}'_{n}) \Delta t_n \\
 \mathbf{u}_{n+1} &=& \mathbf{u}'_{n+1} + C(\mathbf{\dot{u}}'_{n+1} - 
 \mathbf{\dot{u}}'_{n}) \Delta t_n \\
 \mathbf{T}_{n+1} &=& \mathbf{T}'_{n+1} + C(\mathbf{\dot{T}}'_{n+1} - 
 \mathbf{\dot{T}}'_{n}) \Delta t_n
\end{eqnarray*}

\noindent where  $B=1/2$ is required to obtain  accurate velocities to
second order and the values of  $A$ and $C$ are somewhat arbitrary. We
adopt  $A=1/3$ and  $C=1/2$.   Finally, timesteps  ($\Delta t_n$)  are
determined  comparing   the  local  sound  velocity   with  the  local
acceleration and  imposing that  none of the  SPH particles  travels a
distance  larger than  its  corresponding smoothing  length. Also  the
change in  the internal energy is  taken into account  to restrict the
timesteps.   In particular,  timesteps are  computed in  the following
way:
\begin{eqnarray*}
 \Delta t^{n+1} &=& \min_i \Delta t_i^{n+1} \\
 \Delta t^{n+1}_i &=& \min \left( f_h,\sqrt{\frac{h_i^n}{|\mathbf{a}_i^n|}},
f_h\frac{h_i^n}{v_i^{{\rm sig},n}},f_u\Delta t^n\frac{u_i^{n-1}}{u_i^n-u_i^{n-1}} 
\right)
\end{eqnarray*}
with $f_h=0.5$ and $f_u=0.3$. Adopting all these prescriptions, energy
is conserved at the level of 1\%, and angular momentum at the level of
$10^{-3}$\% in all simulations.

The  thermodynamical  properties  of  matter are  computed  using  the
Helmhotz equation of state \citep{TiSw00}. The nuclear network adopted
here incorporates 14 nuclei: He, C, O,  Ne, Mg, Si, S, Ar, Ca, Ti, Cr,
Fe,  Ni and  Zn. The  reactions  considered are  captures of  $\alpha$
particles, and the  associated back reactions (photo-disintegrations),
the fusion  of two C nuclei and  the reaction between C  and O nuclei.
All the thermonuclear  reaction rates are taken from  the REACLIB data
base  \citep{Cy10}.  The screening  factors adopted  in this  work are
those  of  \cite{ItTo79}.   The  nuclear energy  release  is  computed
independently of the dynamical  evolution with much shorter timesteps,
assuming that the dynamical variables  do not change during these time
steps.     Nevertheless,   when   photo-disintegrations    occur   the
temperatures  and  the  specific   heat  are  updated,  following  the
procedure detailed  in \cite{RaSc10}. Nuclear  abundances are obtained
by means  of an iterative pseudo-Gaussian  elimination technique based
on a two-step linearization procedure \citep{Wago69}.

\section{Initial conditions}
\label{ini}

\begin{table*}
\caption{Kinematical  properties  of  the  simulations  reported  here
  involving a $0.8\, M_{\sun}$ white dwarf.}
\label{orbits-CO}
\begin{center}
\small
\begin{tabular}{ccccccccccc}
\hline 
\hline 
\noalign{\smallskip}
 Run & $M_1+M_2$    & Outcome & Detonation & Ejection & $E$             & $L$                      & $r_{\rm max}$ & $r_{\rm min}$     & $\varepsilon$     & $\beta$ \\
     & ($M_{\sun}$) &         &            &            & ($10^{48}$~erg) & ($10^{50}$~erg~s$^{-1}$) & ($R_{\sun}$)  & ($R_{\sun}$)      &                   & \\[0.6ex]
\noalign{\smallskip}
\hline 
\hline
\multicolumn{9}{l}{$v_{\rm ini} = 75 \;\text{km/s} \quad \quad \Delta y = 0.4 \;R_{\sun}$} \\
\hline
\noalign{\smallskip}
1  & 0.8+0.6 & DC & Yes & No & $-4.12$ &  2.80 & 4.48$\times 10^{-1}$ & 6.72$\times 10^{-3}$ & 0.970 & 2.83 \\
2  & 0.8+0.8 & DC & Yes & No & $-5.49$ &  3.28 & 4.49$\times 10^{-1}$ & 5.88$\times 10^{-3}$ & 0.974 & 3.06 \\
3  & 1.0+0.8 & DC & Yes &  2 & $-6.88$ &  3.64 & 4.49$\times 10^{-1}$ & 5.21$\times 10^{-3}$ & 0.977 & 3.07 \\
4  & 1.2+0.8 & DC & Yes &  1 & $-8.26$ &  3.93 & 4.48$\times 10^{-1}$ & 4.70$\times 10^{-3}$ & 0.980 & 2.98 \\
\hline 
\multicolumn{9}{l}{$v_{\rm ini} = 100 \;\text{km/s} \quad \quad \Delta y = 0.3 \;R_{\sun}$} \\
\hline
\noalign{\smallskip}
5  & 0.8+0.6 & DC & Yes & No & $-5.05$ &  2.81 & 3.64$\times 10^{-1}$ & 6.76$\times 10^{-3}$ & 0.964 & 2.81 \\
6  & 0.8+0.8 & DC & Yes & No & $-6.76$ &  3.28 & 3.63$\times 10^{-1}$ & 5.90$\times 10^{-3}$ & 0.968 & 3.05 \\
7  & 1.0+0.8 & DC & Yes &  2 & $-8.47$ &  3.64 & 3.63$\times 10^{-1}$ & 5.23$\times 10^{-3}$ & 0.972 & 3.06 \\
8  & 1.2+0.8 & DC & Yes &  1 & $-10.1$ &  3.93 & 3.63$\times 10^{-1}$ & 4.71$\times 10^{-3}$ & 0.974 & 2.97 \\
\hline 
\multicolumn{9}{l}{$v_{\rm ini} = 100 \;\text{km/s} \quad \quad \Delta y = 0.4 \;R_{\sun}$} \\
\hline
\noalign{\smallskip}
9  & 0.8+0.6 & LC & Yes & No & $-4.06$ &  3.74 & 4.49$\times 10^{-1}$ & 1.21$\times 10^{-2}$ & 0.948 & 1.57 \\
10 & 0.8+0.8 & LC & Yes & No & $-5.42$ &  4.37 & 4.50$\times 10^{-1}$ & 1.05$\times 10^{-2}$ & 0.954 & 1.71 \\
11 & 1.0+0.8 & LC & Yes & No & $-6.80$ &  4.85 & 4.50$\times 10^{-1}$ & 9.35$\times 10^{-3}$ & 0.959 & 1.71 \\
12 & 1.2+0.8 & LC & Yes & No & $-8.18$ &  5.24 & 4.49$\times 10^{-1}$ & 8.42$\times 10^{-3}$ & 0.963 & 1.66 \\
\hline 
\multicolumn{9}{l}{$v_{\rm ini} = 150 \;\text{km/s} \quad \quad \Delta y = 0.3 \;R_{\sun}$} \\
\hline
\noalign{\smallskip}
13 & 0.8+0.6 & LC &  No & No & $-4.88$ &  4.21 & 3.68$\times 10^{-1}$ & 1.56$\times 10^{-2}$ & 0.919 & 1.22 \\
14 & 0.8+0.8 & LC & Yes & No & $-6.56$ &  4.91 & 3.67$\times 10^{-1}$ & 1.35$\times 10^{-2}$ & 0.929 & 1.33 \\ 
15 & 1.0+0.8 & LC & Yes & No & $-8.25$ &  5.46 & 3.66$\times 10^{-1}$ & 1.20$\times 10^{-2}$ & 0.937 & 1.33 \\
16 & 1.2+0.8 & LC & Yes & No & $-9.95$ &  5.90 & 3.66$\times 10^{-1}$ & 1.08$\times 10^{-2}$ & 0.943 & 1.30 \\
\hline 
\multicolumn{9}{l}{$v_{\rm ini} = 150 \;\text{km/s} \quad \quad \Delta y = 0.4 \;R_{\sun}$} \\
\hline
\noalign{\smallskip}
17 & 0.8+0.6 & LC &  No & No & $-3.89$ &  5.60 & 4.53$\times 10^{-1}$ & 2.81$\times 10^{-2}$ & 0.883 & 0.68 \\
18 & 0.8+0.8 & LC &  No & No & $-5.22$ &  6.55 & 4.54$\times 10^{-1}$ & 2.45$\times 10^{-2}$ & 0.898 & 0.73 \\
19 & 1.0+0.8 & LC &  No & No & $-6.57$ &  7.28 & 4.53$\times 10^{-1}$ & 2.16$\times 10^{-2}$ & 0.909 & 0.74 \\
20 & 1.2+0.8 & LC & Yes & No & $-7.93$ &  7.86 & 4.52$\times 10^{-1}$ & 1.94$\times 10^{-2}$ & 0.918 & 0.72 \\
\hline 
\multicolumn{9}{l}{$v_{\rm ini} = 200 \;\text{km/s} \quad \quad \Delta y = 0.3 \;R_{\sun}$} \\
\hline
\noalign{\smallskip}
21& 0.8+0.6 & LC  &  No & No & $-4.64$ &  5.62 & 3.75$\times 10^{-1}$ & 2.86$\times 10^{-2}$ & 0.858 & 0.66 \\
22 & 0.8+0.8 & LC &  No & No & $-6.28$ &  6.55 & 3.73$\times 10^{-1}$ & 2.48$\times 10^{-2}$ & 0.875 & 0.73 \\
23 & 1.0+0.8 & LC &  No & No & $-7.94$ &  7.28 & 3.72$\times 10^{-1}$ & 2.18$\times 10^{-2}$ & 0.889 & 0.73 \\
24 & 1.2+0.8 & LC & Yes & No & $-9.61$ &  7.86 & 3.70$\times 10^{-1}$ & 1.96$\times 10^{-2}$ & 0.900 & 0.71 \\
\hline 
\multicolumn{9}{l}{$v_{\rm ini} = 200 \;\text{km/s} \quad \quad \Delta y = 0.4 \;R_{\sun}$} \\
\hline
\noalign{\smallskip}
25 & 0.8+0.6 & O  &  No & No & $-3.65$ &  7.47 & 4.61$\times 10^{-1}$ & 5.23$\times 10^{-2}$ & 0.796 & 0.36 \\
26 & 0.8+0.8 & O  &  No & No & $-4.94$ &  8.74 & 4.60$\times 10^{-1}$ & 4.54$\times 10^{-2}$ & 0.820 & 0.40 \\   
27 & 1.0+0.8 & O  &  No & No & $-6.26$ &  9.70 & 4.58$\times 10^{-1}$ & 3.99$\times 10^{-2}$ & 0.840 & 0.40 \\
28 & 1.2+0.8 & O  &  No & No & $-7.60$ & 10.50 & 4.57$\times 10^{-1}$ & 3.56$\times 10^{-2}$ & 0.855 & 0.39 \\
\hline
\noalign{\smallskip}
\multicolumn{9}{l}{$v_{\rm ini} = 300 \;\text{km/s} \quad \quad \Delta y = 0.3 \;R_{\sun}$} \\
\hline
\noalign{\smallskip}
29 & 0.8+0.6 &  O &  No & No & $-3.97$ &  8.39 & 4.02$\times 10^{-1}$ & 6.96$\times 10^{-2}$ & 0.705 & 0.27 \\
30 & 0.8+0.8 &  O &  No & No & $-5.47$ &  9.32 & 3.96$\times 10^{-1}$ & 6.02$\times 10^{-2}$ & 0.736 & 0.30 \\
31 & 1.0+0.8 &  O &  No & No & $-7.04$ & 10.90 & 3.90$\times 10^{-1}$ & 5.27$\times 10^{-2}$ & 0.762 & 0.30 \\
32 & 1.2+0.8 &  O &  No & No & $-8.64$ & 11.80 & 3.86$\times 10^{-1}$ & 4.69$\times 10^{-2}$ & 0.783 & 0.30 \\
\hline
\multicolumn{9}{l}{$v_{\rm ini} = 300 \;\text{km/s} \quad \quad \Delta y = 0.4 \;R_{\sun}$} \\
\hline
\noalign{\smallskip}
33 & 0.8+0.6 &  O &  No & No & $-2.94$ & 11.20 & 5.01$\times 10^{-1}$ & 1.35$\times 10^{-1}$  & 0.576 & 0.14 \\
34 & 0.8+0.8 &  O &  No & No & $-4.13$ & 13.10 & 4.89$\times 10^{-1}$ & 1.15$\times 10^{-1}$  & 0.620 & 0.16 \\ 
35 & 1.0+0.8 &  O &  No & No & $-5.37$ & 14.62 & 4.82$\times 10^{-1}$ & 9.97$\times 10^{-2}$  & 0.657 & 0.16 \\
36 & 1.2+0.8 &  O &  No & No & $-6.63$ & 15.71 & 4.76$\times 10^{-1}$ & 8.81$\times 10^{-2}$  & 0.688 & 0.16 \\
\noalign{\smallskip}
\hline
\hline
\end{tabular}
\end{center}
\end{table*}

\begin{table*}
\caption{Kinematical  properties  of  the  simulations  reported  here
  involving a $0.4\, M_{\sun}$ white dwarf.}
\label{orbits-He}
\begin{center}
\small
\begin{tabular}{ccccccccccc}
\hline 
\hline 
\noalign{\smallskip}
 Run & $M_1+M_2$    & Outcome & Detonation & Ejection & $E$             & $L$                      & $r_{\rm max}$ & $r_{\rm min}$     & $\varepsilon$     & $\beta$ \\
     & ($M_{\sun}$) &         &            &            & ($10^{48}$~erg) & ($10^{50}$~erg~s$^{-1}$) & ($R_{\sun}$)  & ($R_{\sun}$)      &                   & \\[0.6ex]
\noalign{\smallskip}
\hline 
\hline
\multicolumn{9}{l}{$v_{\rm ini} = 75 \;\text{km/s} \quad \quad \Delta y = 0.3 \;R_{\sun}$} \\
\hline
\noalign{\smallskip}
37 & 0.2+0.4 & DC & Yes & 1 & $-0.84$ & 0.82 & 3.65$\times 10^{-1}$ & 8.94$\times 10^{-3}$ & 0.952 & 3.97 \\
38 & 0.4+0.4 & DC & Yes & 2 & $-1.69$ & 1.23 & 3.64$\times 10^{-1}$ & 6.65$\times 10^{-3}$ & 0.964 & 4.21 \\
\hline
\multicolumn{9}{l}{$v_{\rm ini} = 75 \;\text{km/s} \quad \quad \Delta y = 0.4 \;R_{\sun}$} \\
\hline
\noalign{\smallskip}
39 & 0.2+0.4 & LC & Yes & No & $-0.67$ &  1.09 & 4.51$\times 10^{-1}$ & 1.60$\times 10^{-2}$ & 0.931 & 2.21 \\
40 & 0.4+0.4 & DC & Yes &  2 & $-1.35$ &  1.64 & 4.50$\times 10^{-1}$ & 1.19$\times 10^{-2}$ & 0.948 & 2.35 \\
41 & 0.8+0.4 & DC & Yes &  1 & $-2.73$ &  2.18 & 4.49$\times 10^{-1}$ & 7.85$\times 10^{-3}$ & 0.966 & 2.93 \\
42 & 1.2+0.4 & DC & Yes &  1 & $-4.12$ &  2.45 & 4.48$\times 10^{-1}$ & 5.84$\times 10^{-3}$ & 0.974 & 3.26 \\ 
\hline
\multicolumn{9}{l}{$v_{\rm ini} = 100 \;\text{km/s} \quad \quad \Delta y = 0.3 \;R_{\sun}$} \\
\hline
\noalign{\smallskip}
43 & 0.2+0.4 & LC & Yes & No & $-0.81$ &  1.09 & 3.68$\times 10^{-1}$ & 1.62$\times 10^{-2}$ & 0.916 & 2.19 \\
44 & 0.4+0.4 & DC & Yes &  2 & $-1.65$ &  1.64 & 3.66$\times 10^{-1}$ & 1.20$\times 10^{-2}$ & 0.937 & 2.34 \\
45 & 0.8+0.4 & DC & Yes &  1 & $-3.35$ &  2.18 & 3.64$\times 10^{-1}$ & 7.88$\times 10^{-3}$ & 0.958 & 2.92 \\
46 & 1.2+0.4 & DC & Yes &  1 & $-5.07$ &  2.45 & 3.63$\times 10^{-1}$ & 5.88$\times 10^{-3}$ & 0.968 & 3.23 \\
\hline
\multicolumn{9}{l}{$v_{\rm ini} = 100 \;\text{km/s} \quad \quad \Delta y = 0.4 \;R_{\sun}$} \\
\hline
\noalign{\smallskip}
47 & 0.2+0.4 & LC &  No & No & $-0.64$ &  1.46 & 4.55$\times 10^{-1}$ & 2.93$\times 10^{-2}$ & 0.879 & 1.21 \\
48 & 0.4+0.4 & LC & Yes & No & $-1.31$ &  2.18 & 4.53$\times 10^{-1}$ & 2.16$\times 10^{-2}$ & 0.909 & 1.29 \\
49 & 0.8+0.4 & DC & Yes &  1 & $-2.68$ &  2.91 & 4.51$\times 10^{-1}$ & 1.42$\times 10^{-2}$ & 0.939 & 1.62 \\
50 & 1.2+0.4 & DC & Yes &  1 & $-4.07$ &  3.27 & 4.50$\times 10^{-1}$ & 1.05$\times 10^{-2}$ & 0.954 & 1.80 \\
\hline
\multicolumn{9}{l}{$v_{\rm ini} = 150 \;\text{km/s} \quad \quad \Delta y = 0.3 \;R_{\sun}$} \\
\hline
\noalign{\smallskip}
51 & 0.2+0.4 & LC &  No & No & $-0.74$ &  1.64 & 3.81$\times 10^{-1}$ & 3.84$\times 10^{-2}$ & 0.818 & 0.92 \\
52 & 0.4+0.4 & LC & Yes & No & $-1.55$ &  2.46 & 3.75$\times 10^{-1}$ & 2.81$\times 10^{-2}$ & 0.861 & 1.00 \\
53 & 0.8+0.4 & DC & Yes &  1 & $-3.22$ &  3.27 & 3.69$\times 10^{-1}$ & 1.83$\times 10^{-2}$ & 0.906 & 1.26 \\
54 & 1.2+0.4 & DC & Yes &  1 & $-4.92$ &  3.68 & 3.67$\times 10^{-1}$ & 1.35$\times 10^{-2}$ & 0.929 & 1.41 \\
\hline
\multicolumn{9}{l}{$v_{\rm ini} = 150 \;\text{km/s} \quad \quad \Delta y = 0.4 \;R_{\sun}$} \\
\hline
\noalign{\smallskip}
55 & 0.2+0.4 & O  &  No & No & $-0.58$ &  2.18 & 4.70$\times 10^{-1}$ & 7.14$\times 10^{-2}$ & 0.736 & 0.50 \\
56 & 0.4+0.4 & O  &  No & No & $-1.21$ &  3.28 & 4.62$\times 10^{-1}$ & 5.16$\times 10^{-2}$ & 0.799 & 0.54 \\
57 & 0.8+0.4 & LC & Yes & No & $-2.55$ &  4.37 & 4.56$\times 10^{-1}$ & 3.30$\times 10^{-2}$ & 0.864 & 0.70 \\
58 & 1.2+0.4 & LC & Yes & No & $-3.91$ &  4.91 & 4.54$\times 10^{-1}$ & 2.44$\times 10^{-2}$ & 0.898 & 0.78 \\
\hline
\multicolumn{9}{l}{$v_{\rm ini} = 200 \;\text{km/s} \quad \quad \Delta y = 0.3 \;R_{\sun}$} \\
\hline
\noalign{\smallskip}
59 & 0.2+0.4 & O  &  No & No & $-0.65$ &  2.18 & 4.07$\times 10^{-1}$ & 7.31$\times 10^{-2}$ & 0.695 & 0.49 \\
60 & 0.4+0.4 & O  &  No & No & $-1.41$ &  3.28 & 3.90$\times 10^{-1}$ & 5.27$\times 10^{-2}$ & 0.762 & 0.53 \\
61 & 0.8+0.4 & LC & Yes & No & $-3.03$ &  4.36 & 3.78$\times 10^{-1}$ & 3.36$\times 10^{-2}$ & 0.837 & 0.68 \\
62 & 1.2+0.4 & LC & Yes & No & $-4.71$ &  4.91 & 3.73$\times 10^{-1}$ & 2.47$\times 10^{-2}$ & 0.876 & 0.77 \\
\hline
\multicolumn{9}{l}{$v_{\rm ini} = 200 \;\text{km/s} \quad \quad \Delta y = 0.4 \;R_{\sun}$} \\
\hline
\noalign{\smallskip}
63 & 0.2+0.4 & O  &  No & No & $-0.48$ &  2.91 & 5.05$\times 10^{-1}$ & 1.41$\times 10^{-1}$ & 0.564 & 0.25 \\  
64 & 0.4+0.4 & O  &  No & No & $-1.07$ &  4.37 & 4.82$\times 10^{-1}$ & 9.97$\times 10^{-2}$ & 0.657 & 0.28 \\  
65 & 0.8+0.4 & O  &  No & No & $-2.36$ &  5.82 & 4.66$\times 10^{-1}$ & 6.23$\times 10^{-2}$ & 0.764 & 0.37 \\
66 & 1.2+0.4 & LC & Yes & No & $-3.70$ &  6.55 & 4.60$\times 10^{-1}$ & 4.53$\times 10^{-2}$ & 0.821 & 0.42 \\
\hline
\multicolumn{9}{l}{$v_{\rm ini} = 300 \;\text{km/s} \quad \quad \Delta y = 0.3 \;R_{\sun}$} \\
\hline
\noalign{\smallskip}
67 & 0.4+0.4 &  O &  No & No & $-1.01$ &  4.91 & 4.87$\times 10^{-1}$ & 1.33$\times 10^{-1}$ & 0.571 & 0.14 \\
68 & 0.8+0.4 &  O &  No & No & $-2.50$ &  6.55 & 4.17$\times 10^{-1}$ & 8.35$\times 10^{-2}$ & 0.666 & 0.28 \\
69 & 1.2+0.4 &  O &  No & No & $-4.11$ &  7.36 & 3.96$\times 10^{-1}$ & 6.01$\times 10^{-2}$ & 0.737 & 0.32 \\
\hline
\multicolumn{9}{l}{$v_{\rm ini} = 300 \;\text{km/s} \quad \quad \Delta y = 0.4 \;R_{\sun}$} \\
\hline
\noalign{\smallskip}
70 & 0.8+0.4 &  O &  No & No & $-1.82$ &  8.74 & 5.22$\times 10^{-1}$ & 1.63$\times 10^{-1}$  & 0.525 & 0.14 \\
71 & 1.2+0.4 &  O &  No & No & $-3.10$ &  9.82 & 4.90$\times 10^{-1}$ & 1.14$\times 10^{-1}$  & 0.621 & 0.17 \\
\noalign{\smallskip}
\hline
\hline
\end{tabular}
\end{center}
\end{table*}

We  have  relaxed  five  different  white  dwarf  models  with  masses
$M_1=0.4\, M_{\sun},  0.6\, M_{\sun}, 0.8\,  M_{\sun}, 1.0\, M_{\sun}$
and $1.2\, M_{\sun}$.  The chemical composition of all white dwarfs is
a mixture  of 40\% carbon  and 60\% oxygen  (by mass), except  for the
lightest one --- which is made of pure helium --- and the heaviest one
--- which is  made of a  mixture of 80\%  of oxygen and 20\%  of neon,
also by mass.  The intervening stars were relaxed separately to obtain
accurate  equilibrium  initial configurations,  using  $\sim 2  \times
10^5$  particles for  each star.   This resolution  is high  enough to
provide  accurate results  and low  enough to  run a  large  number of
simulations in  a reasonable period of time.   The initial temperature
of our isothermal white dwarf  configurations is $\sim 10^7$~K.  As in
\cite{PaEn10},   the    white   dwarfs   were    assumed   to   rotate
counterclockwise  as   rigid  bodies  \citep{Cha09}   with  rotational
velocities $\omega\simeq 7\times 10^{-5}$~rad/s --- a typical rotation
velocity of  field white dwarfs  \citep{BK05}.  This rotation  rate is
nevertheless  irrelevant  to  the  dynamics of  the  close  encounters
studied in this  paper. In fact, the spin  rates of post-capture white
dwarfs  could  be  larger,  as  the  capture  is  dominated  by  tidal
dissipation, which  would lead to spin-up  of at least  the lower mass
white  dwarf,  and  of  both   stars  if  the  masses  are  comparable
\citep{PT77}.

\begin{figure}
\begin{center}
\includegraphics[width=7cm]{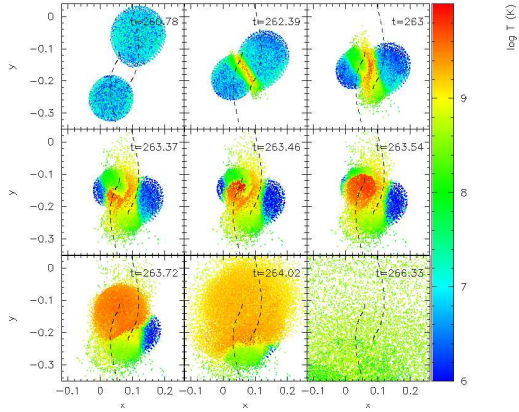}
\caption{Time  evolution  of  one  of  the simulations  in  which  the
  dynamical interaction of two  white dwarfs results in the disruption
  of both  stars.  In particular,  this simulation corresponds  to the
  case in which the interacting stars have masses $0.8\, M_{\sun}$ and
  $1.0\,   M_{\sun}$,  whereas   the  initial   velocity   is  $v_{\rm
  ini}=100$~km/s and  the initial distance is  $\Delta y=0.3 R_{\sun}$
  --- that is, run number 7 in Table~\ref{orbits-CO}.  The temperature
  of each SPH particle is also shown, expressed in K.  The $x$ and $y$
  axes are in units of  $0.1\, R_{\sun}$.  The dashed lines correspond
  to  the trajectories  of  the  center of  mass  of each  intervening
  star. Only  1 out of 10  particles has been  represented.  Times (in
  seconds)  since the  beginning of  the simulation  are shown  in the
  right  upper corner  of each  panel.  These  figures have  been done
  using the  visualization tool SPLASH  \citep{Price07}. [Color figure
  only available in the electronic version of the article].}
\label{CO-disrupted}
\end{center}
\end{figure}

We fixed  the initial distance  between the stars along  the $x$-axis,
$\Delta x=0.2\, R_{\sun}$, and allowed  the initial distance along the
$y$-axis  to   vary  between   $\Delta  y=0.3\,R_{\sun}$   and  $0.4\,
R_{\sun}$.   Under these  conditions  the tidal  deformations of  both
white dwarfs are negligible at the beginning of the simulation and the
approximation of spherical symmetry is  valid. The initial velocity of
each star  was set to  ${\mathbf v}_{\rm ini}=(\pm  v_{\rm ini},0,0)$,
with  $v_{\rm ini}$  ranging from  75 to  300~km/s, which  are typical
values for  which the interaction ends  up in a collision.   With this
setting the  initial coordinates of  the two intervening  white dwarfs
are $(\Delta x/2,-\Delta y/2,0)$ and $(-\Delta x/2, \Delta y/2,0)$ and
the  relative velocity  is $2v_{\rm  ini}$.  Note  that these  initial
conditions lead  in all  cases to negative  energies, which  result in
initial  elliptical  trajectories, although  some  of  them have  high
eccentricities --- see below.  That  is, the interactions studied here
correspond to  a post-capture  scenario. For a  detailed study  of the
gravitational  capture mechanisms  see, for  example, \cite{PT77}  and
\cite{LO86}. Additionally, we  note that in order for a  pair of stars
to  become bound  after a  close encounter,  some kind  of dissipation
mechanism  must  be involved,  like  a  third body  tidal  interaction
\citep{SH02},  or the  excitation of  stellar pulsations  by means  of
tidal interaction \citep{Fea75}.

\begin{figure}
\begin{center}
\includegraphics[width=7cm]{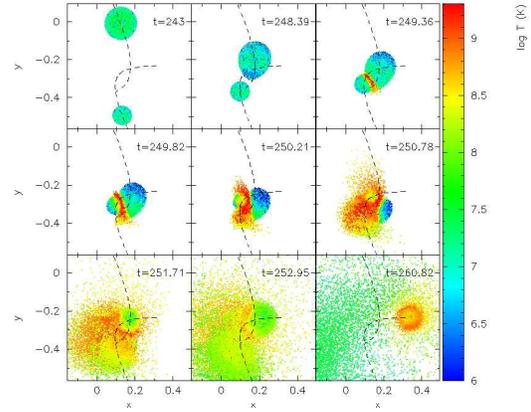}
\caption{Same as Fig.~\ref{CO-disrupted} for one of the simulations in
  which  the dynamical interaction  results in  the disruption  of the
  less massive  star, while the  more massive white dwarf  retains its
  identity. This specific simulation  corresponds to the case in which
  the  initial  velocity is  $v_{\rm  ini}=100$~km/s  and the  initial
  distance  is $y_{\rm  ini}=0.3 R_{\sun}$,  while the  masses  of the
  colliding white dwarfs are $0.8$ and $1.2\, M_{\sun}$, respectively,
  corresponding  to  simulation  number  8  in  Table~\ref{orbits-CO}.
  [Color  figure  only available  in  the  electronic  version of  the
  article].}
\label{ONe-disrupted}
\end{center}
\end{figure}

\section{Outcomes of the interactions}
\label{results}

\subsection{Time evolution}

In most  simulations the time  evolution of the  interactions computed
here  is the  same found  in  our previous  paper \citep{PaEn10}.   In
particular,  after  tidal interaction,  the  intervening white  dwarfs
either form  an eccentric  binary or collide.   In particular,  if the
intervening  stars  get  sufficiently  close at  periastron  and  mass
transfer begins, a stellar merger occurs.  In this case, two different
behaviors   can  be   clearly   distinguished.   If   more  than   one
mass-transfer episode occurs  before the stellar merger, we  name it a
lateral collision (LC).  Else, if just one mass transfer happens, then
we call  the interaction a  direct collision (DC).  Otherwise,  if the
binary system  survives without transferring mass,  an eccentric orbit
will be the outcome (O) of the interaction.

\begin{figure}
\begin{center}
\includegraphics[width=7cm]{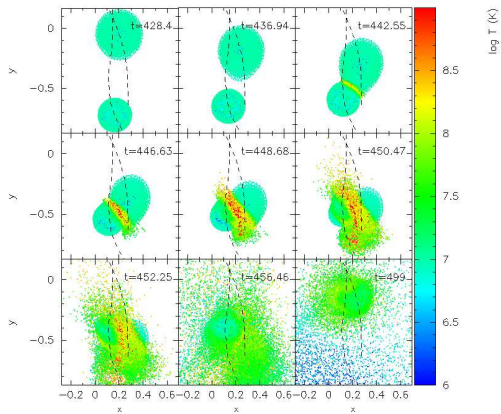}
\caption{Same as  Fig.~\ref{CO-disrupted} for the  simulation in which
  two helium  white dwarfs  are involved. In  this case  the dynamical
  interaction results in the tidal disruption of an extremely low mass
  white dwarf of mass $0.2\, M_{\sun}$ in the gravitational field of a
  $0.4\, M_{\sun}$ helium white  dwarf. The initial conditions of this
  specific simulation  are $v_{\rm ini}=75$~km/s  and $y_{\rm ini}=0.3
  R_{\sun}$, corresponding to  run number 37 in Table~\ref{orbits-He}.
  [Color  figure  only available  in  the  electronic  version of  the
  article].}
\label{He-disrupted}
\end{center}
\end{figure}

Also, for most of the cases  studied here in which a collision occurs,
the resulting  remnants left behind  by the dynamical  interaction are
very  similar to  those  found in  our  previous work  \citep{PaEn10}.
However, there are  a few cases in which the  interaction is so strong
that  the material of  the lightest  white dwarf  is ejected  from the
system.  Finally,  there are as  well some interactions in  which both
stars are totally disrupted.  This occurs as a consequence of the very
high  temperatures attained  in  the contact  region  during the  most
violent phase of the  dynamical interaction.  Specifically, in all the
simulations in which one or  both stars are disrupted and the material
is  ejected from  the system  the temperatures  and  densities reached
during the interaction are high  enough to drive a detonation.  If the
material of the disrupted less massive star is a mixture of carbon and
oxygen this occurs when the temperature is larger than $\sim 2.5\times
10^9$~K  and   the  density  is   above  $2.0\times  10^6$~g~cm$^{-3}$
\citep{Seitenzahl09, Pakmor11}.  When the  less massive white dwarf is
made of helium we consider that a detonation is likely to develop when
the   nuclear   timescale  is   shorter   than   the  dynamical   one.
Nevertheless, we  emphasize that these are  only necessary conditions,
since whether  a detonation develops or  not depends as  well on other
factors, like the temperature and density gradients.
 
We find that  in most of the simulations in which  the material of the
disrupted  low-mass white  dwarf reaches  high temperatures  and large
densities  the regions  in which  a  detonation is  likely to  develop
comprise a small number of particles and degeneracy is rapidly lifted.
Consequently, in  these cases the result of  the dynamical interaction
is  not a powerful  thermonuclear explosion,  leading to  a supernova.
However, there  are a few runs  in which the number  of particles that
reach detonation conditions is large enough to ensure that a supernova
occurs.  This  happens, for instance, in  the case in  which two heavy
carbon-oxygen  white  dwarfs of  masses  $0.8\,  M_{\sun}$ and  $1.0\,
M_{\sun}$ interact.  The time evolution  of this system is depicted in
Fig.~\ref{CO-disrupted}.   There are  other  cases in  which only  the
material  of the  less  massive  white dwarf  is  ejected after  being
tidally disrupted by the more massive one.  This occurs, for instance,
when the  primary is a very  compact oxygen-neon white  dwarf.  Due to
the  very small  radius  of the  more  massive white  dwarf a  sizable
fraction of the less massive white  dwarf is not accreted as it occurs
when  two  white  dwarfs  with  smaller mass  contrast  interact  but,
instead, in this  case the oxygen-neon white dwarf  is barely affected
by the interaction,  while the material of the  disrupted less massive
star bounces on the surface of  the primary and is ejected at somewhat
large     velocities,     of     the     order     of     $10^4$~km/s.
Fig.~\ref{ONe-disrupted} shows an example of the temporal evolution in
these cases. Specifically, this  figure displays the time evolution in
the case in  which a $0.8\, M_{\sun}$ carbon-oxygen  white dwarf and a
$1.2\ M_{\sun}$ oxygen-neon white  dwarf interact.  Finally, there are
other simulations in  which both stars are disrupted  as well although
there  is not  a  large mass  contrast.   This occurs  mainly for  the
simulations      involving      two      helium     white      dwarfs.
Fig.~\ref{He-disrupted} illustrates the case in which an extremely low
mass  white dwarf  of mass  $0.2\,  M_{\sun}$ white  dwarf is  tidally
disrupted by another helium white  dwarf of mass $0.4\, M_{\sun}$, and
its material is  ejected from the system.  Finally,  it is interesting
to note  as well that in  all these simulations the  shocked region is
well  resolved by  our simulations  ---  see Figs.~\ref{CO-disrupted},
\ref{ONe-disrupted}, and \ref{He-disrupted}.

\subsection{Overview of the simulations}

\begin{figure*}
\vspace{8.1cm}
\includegraphics{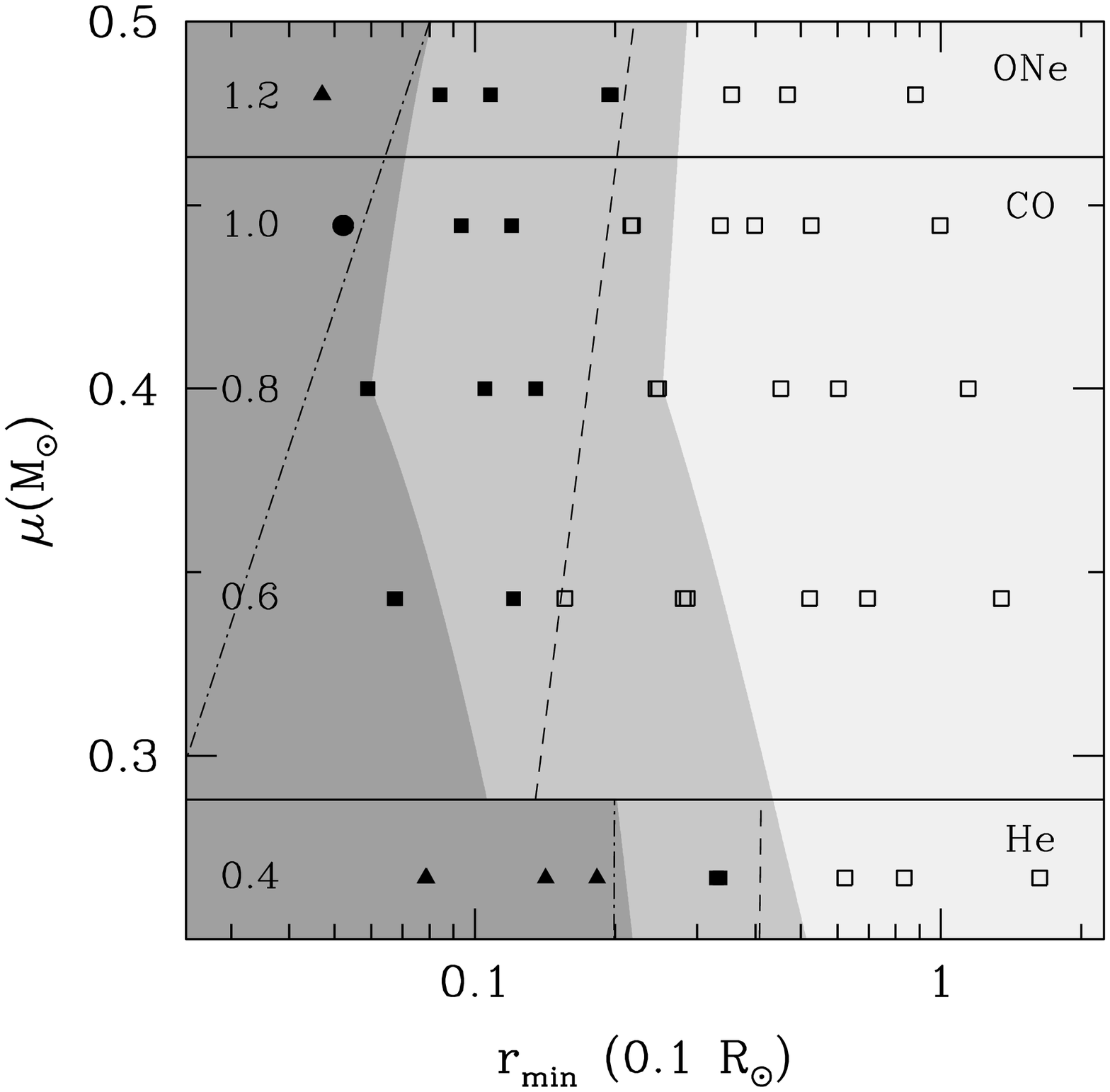}
\includegraphics{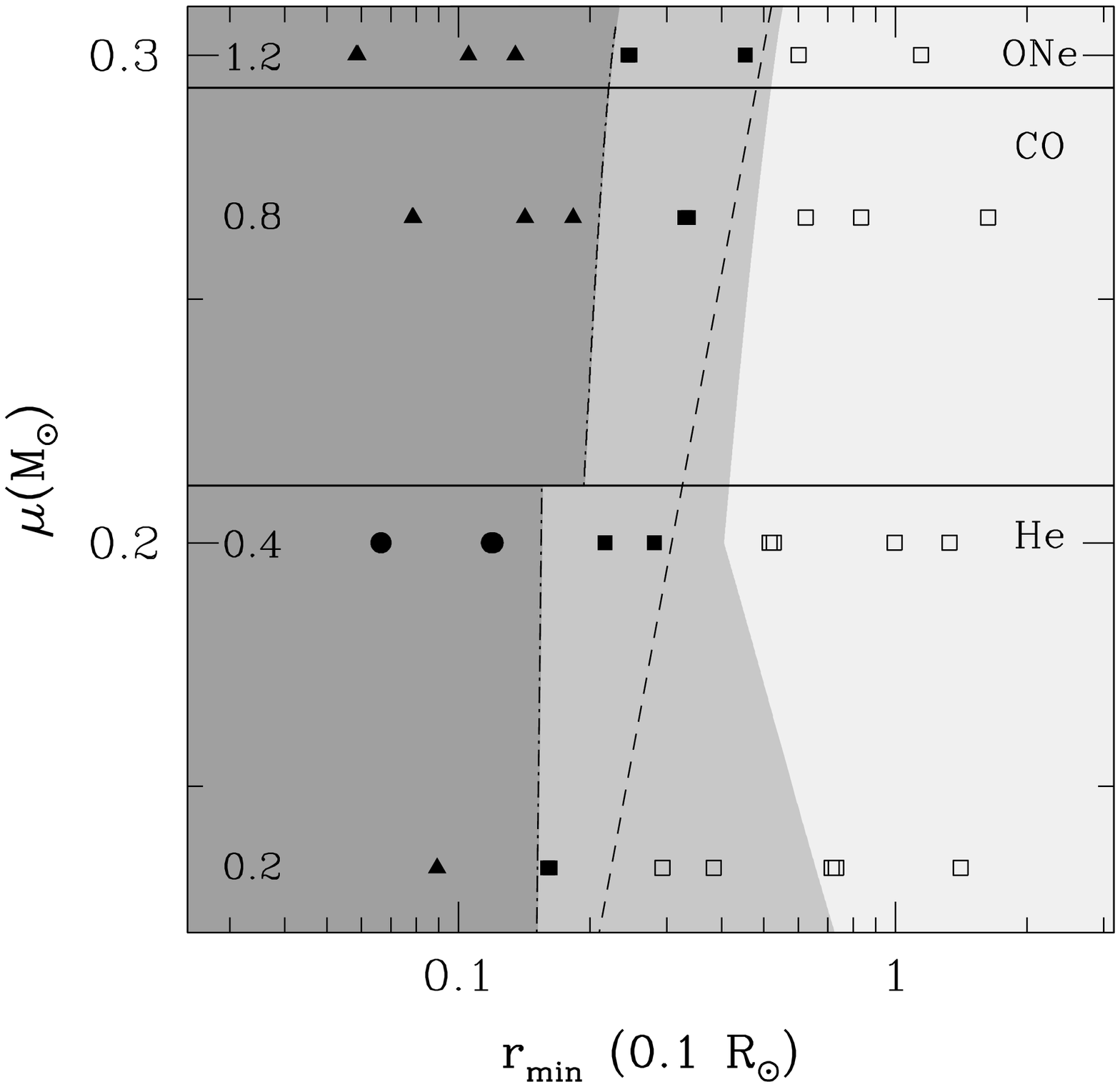}
\caption{Outcomes of  the simulations performed here  presented in the
  plane defined by  the reduced mass of the  system and the periastron
  distance. The left panel displays the outcomes of the simulations in
  which  a $0.8 \,  M_{\sun}$ carbon-oxygen  white dwarf  is involved,
  whilst  the right  panel shows  the outcomes  of the  simulations in
  which a  $0.4 \, M_{\sun}$ helium  white dwarf is  used.  The shaded
  areas indicate  the regions for  which the three  different outcomes
  occur, as  explained in the main  text.  Each row  of simulations is
  also labelled  with the mass  of the second interacting  white dwarf
  corresponding to  the indicated reduced mass.   The horizontal solid
  lines separate the regions  of helium, carbon-oxygen and oxygen-neon
  white   dwarfs,   respectively.   Hollow  squares   indicate   those
  simulations in which no  detonation occurs, filled squares represent
  those simulations in which the  conditions for a detonation are met,
  filled  triangles  correspond  to  those simulations  in  which  the
  material  of  the lightest  white  dwarf  is  ejected, while  filled
  circles  show  the  simulations  in  which  both  white  dwarfs  are
  completely destroyed.  The dashed line separates the region in which
  no  detonations  occur  and  that  in  which  during  the  dynamical
  interaction the  physical conditions for  a detonation to  occur are
  met, while the dotted-dashed line indicates the region for which the
  dynamically  interacting system is  totally or  partially disrupted.
  Those interactions  occuring to  the left of  this line result  in a
  ejection of the material of at least one component of the system.}
\label{plane}
\end{figure*}

Tables~\ref{orbits-CO}  and \ref{orbits-He}  list all  the simulations
performed in this work, grouped as a function of the initial positions
and  velocities.   In   particular,  Table~\ref{orbits-CO}  lists  the
kinematical properties of the  simulations in which a $0.8\, M_{\sun}$
carbon-oxygen     white    dwarf     is    involved,     whereas    in
Table~\ref{orbits-He}  we   display  the  same   information  for  the
simulations in  which a  light-weight ($0.4\, M_{\sun}$)  helium white
dwarf is considered.  Note  that these simulations complement those of
\cite{PaEn10}, in which the interaction  of two white dwarfs of masses
$0.6$  and $0.8\,  M_{\sun}$ was  studied.  Thus,  the present  set of
simulations,   when   complemented   with  those   of   \cite{PaEn10},
encompasses the most plausible range of white dwarf masses.

In  both tables we  list the  masses of  the interacting  white dwarfs
(second column)  and the outcome  of the dynamical  interaction (third
column).  The  fourth column  shows if the  physical conditions  for a
detonation  are met  during the  dynamical  interaction, and  if as  a
consequence of the interaction the  material of the less massive white
dwarf is ejected  (1), or if both stars are  totally disrupted (2) ---
fifth column.  The total energy,  $E$, and the total angular momentum,
$L$,  of the  system  are listed  in  the sixth  and seventh  columns,
respectively.   These  two quantities  have  been  computed using  the
corresponding SPH  prescriptions.  Note as well that  all the energies
of the  systems considered  here are negative  and, thus,  the initial
trajectories are  in all  cases elliptical.  Therefore,  the apoastron
($r_{\rm max}$),  the periastron ($r_{\rm min}$)  and the eccentricity
($\epsilon$)  of the  orbits  are  also specified  in  this table  ---
columns  7, 8  and 9,  respectively.  All  these quantities  have been
computed using the solution of the two-body problem, assuming that the
two  white dwarfs  are  point  masses.  However,  it  is important  to
realize  that  tidal  interactions  subsequently  modify  the  initial
orbits, leading to different outcomes.  Finally, in the last column of
tables \ref{orbits-CO} and \ref{orbits-He}  we also list the so-called
impact parameter, which is defined as
\begin{equation}
\beta=\frac{R_{1}+R_{2}}{r_{\rm min}},
\label{beta}
\end{equation}
This parameter is a good indicator of the strength of the interaction.
In this expression $R_1$ is the radius of the more massive white dwarf
and $R_2$ that of the less massive one.

\subsection{The outcomes of the interactions}

As  previously  said,  our   simulations  result  in  three  different
outcomes,   depending  on   the  adopted   initial   conditions.   Not
surprisingly, we  find that the  most relevant physical  parameter for
discriminating between the three  different outcomes is the periastron
distance,  $r_{\rm  min}$.   In   particular,  we  find  that  as  the
periastron distance decreases, the  outcome of the interaction changes
from the formation  of an eccentric binary to  a lateral collision and
then to a direct one.  This  can be seen in Fig.~\ref{plane}, where we
show the  different outcomes of our  simulations as a  function of the
periastron  distance   and  the  reduced   mass  of  the   system  ---
$\mu=M_1M_2/(M_1+M_2)$ ---  for the case in  which the mass  of one of
the interacting  white dwarfs  is kept fixed  to $0.8\,  M_{\sun}$ ---
left  panel ---  or to  $0.4\, M_{\sun}$  --- right  panel.   In these
panels the  dark grey shaded areas  represent the region  in which the
outcome  of the  interaction is  a direct  collision, the  medium grey
shaded areas show the regions in which lateral collisions occur, while
the light  grey shaded  areas display the  regions in  which eccentric
binaries are formed.  We have  also labelled, for the sake of clarity,
each set of simulations with  the mass of the second interacting white
dwarf --- from  $1.2\, M_{\sun}$ to $0.2\, M_{\sun}$.   In this way it
is better illustrated  how the masses of the  white dwarfs involved in
the interaction affect the resulting outcome.

To  gain insight in  the physics  of the  interaction process  we have
proceeded  as follows.   It could  be naively  expected that  a direct
collision occurs  when at closest  approach the two  intervening white
dwarfs are in contact.  This happens when the minimum distance between
their respective centers  of mass, $r_{\rm min}$, is  smaller than the
sum  of the  unperturbed radii  of the  white dwarfs.   That  is, when
$r_{\rm min} \le R_1+R_2$, where $R_1$ and $R_2$ are the radii of both
white  dwarfs at  sufficiently large  distances. Instead,  our results
show that in a direct  collision the overlap between both white dwarfs
at minimum  distance is substantial  in all cases, otherwise  the less
massive star  survives the first  mass transfer episode and  a lateral
collision is the  outcome of the interaction, and  thus this criterion
is not valid.   There are several reasons for this.   The first one is
that in direct collisions the relative velocities of the SPH particles
are  relatively   large,  or  equivalently   $\beta$  is  sufficiently
large. Thus, in direct collisions  the SPH particles of both stars are
thoroughly  mixed.  The second  reason is  that in  this approximation
$r_{\rm  min}$ is  computed  using the  expression  for point  masses,
whereas in our simulations we  deal with extended bodies. Thus, in our
calculations  the distance  at closest  approach is  larger  than that
obtained using point masses.   Finally, tidal forces also decrease the
periastron of the system.  The  interplay between all these factors is
complex   and,  accordingly,   the  simplistic   criterion  previously
explained has to be modified.  To take into account that in all direct
collisions  substantial overlap  at minimum  distance occurs  we adopt
$r_{\rm min}  \le \lambda  R_1 + R_2$,  where the  parameter $\lambda$
indicates the  degree of overlaping between both  stars.  The division
between  the regions of  direct collisions  and lateral  collisions in
Fig.~\ref{plane} is  best fitted using  $\lambda\approx-0.35$, meaning
that indeed the  overlap has to be relatively large.   We note at this
point that the value of $\lambda$ is the same for those simulations in
which either carbon-oxygen and  oxygen-neon white dwarfs are involved,
but not for the case in  which helium white dwarfs interact, for which
the  conditions  for  a  detonation  to occur  are  rather  different.
Nevertheless, this,  in turn, means that this  reasoning is relatively
robust enough,  as it does  not depend much  on the mass of  the white
dwarf.

We now go one step forward and we try to explain which is the physical
mechanism that makes the difference between those simulations in which
a lateral collision  occurs and those simulations which  end up in the
formation of  an eccentric binary.  It would also be  naively expected
that a lateral  collision occurs when the periastron  distance is such
that  the less massive  white dwarf  fills its  Roche lobe  at closest
approach, and consequently mass transfer from the lightest intervening
white  dwarf  to  the most  massive  one  is  enabled.  To  test  this
possibility  we   have  used   the  usual  analytical   expression  of
\cite{Egg83}:

\begin{equation}
R_{\rm L}/a=\frac{0.49 q^{2/3}}{0.6q^{2/3}+\ln(1+q^{1/3})}
\label{RL}
\end{equation}

\begin{table*}
\caption{Hydrodynamical  results  for   the  simulations  in  which  a
  collision occurs and the resulting system remains bound.}
\label{hydro-nodisrupt}
\begin{center}
\begin{tabular}{ccccccccccc}
\hline
\hline
\noalign{\smallskip}
 Run & Detonation & $M_{\rm WD}$  & $M_{\rm corona}$   & $M_{\rm debris}$ & $M_{\rm ej}$  & $T_{\rm max}$   & $T_{\rm peak}$ &  $R_{\rm corona}$ & $R_{\rm debris}$ & $E_{\rm nuc}$  \\
    & ($M_{\sun}$) &  & ($M_{\sun}$) & ($M_{\sun}$) & ($M_{\sun}$) & (K) & (K) &  ($R_{\sun}$) & ($R_{\sun}$) & (erg) \\
\noalign{\smallskip}
\hline 
\hline
\noalign{\smallskip}
 1& Yes & 0.90 & 0.79 & 0.46 & 3.74$\times 10^{-2}$ & 5.35$\times 10^{8}$ & 4.37$\times 10^{9}$ & 9.64$\times 10^{-3}$ & 0.43 & 1.09$\times 10^{49}$ \\
 2& Yes & 1.12 & ---  & 0.34 & 1.41$\times 10^{-1}$ & 4.69$\times 10^{8}$ & 5.44$\times 10^{9}$ & ---                  & 0.28 & 4.57$\times 10^{49}$ \\
 5& Yes & 0.87 & 0.79 & 0.50 & 3.22$\times 10^{-2}$ & 4.90$\times 10^{8}$ & 4.50$\times 10^{9}$ & 8.86$\times 10^{-3}$ & 0.22 & 9.81$\times 10^{48}$ \\  
 6& Yes & 1.13 & ---  & 0.33 & 1.37$\times 10^{-1}$ & 4.65$\times 10^{8}$ & 5.37$\times 10^{9}$ & ---                  & 0.19 & 4.25$\times 10^{49}$ \\
 9& Yes & 0.91 & 0.72 & 0.48 & 1.39$\times 10^{-2}$ & 4.79$\times 10^{8}$ & 3.03$\times 10^{9}$ & 8.00$\times 10^{-3}$ & 0.21 & 9.93$\times 10^{47}$ \\  
10& Yes & 1.29 & ---  & 0.28 & 2.63$\times 10^{-2}$ & 4.50$\times 10^{8}$ & 4.20$\times 10^{9}$ & ---                  & 0.21 & 2.27$\times 10^{48}$ \\
11& Yes & 1.12 & 0.89 & 0.64 & 4.64$\times 10^{-2}$ & 7.77$\times 10^{8}$ & 4.51$\times 10^{9}$ & 6.93$\times 10^{-3}$ & 0.24 & 1.66$\times 10^{49}$ \\
12& Yes & 1.26 & 0.54 & 0.64 & 1.01$\times 10^{-1}$ & 1.08$\times 10^{9}$ & 4.66$\times 10^{9}$ & 4.98$\times 10^{-3}$ & 0.21 & 3.67$\times 10^{49}$ \\
13&  No & 0.91 & 0.71 & 0.48 & 7.96$\times 10^{-3}$ & 4.49$\times 10^{8}$ & 2.18$\times 10^{9}$ & 8.05$\times 10^{-3}$ & 0.22 & 1.49$\times 10^{46}$ \\
14& Yes & 1.30 & ---  & 0.29 & 8.44$\times 10^{-3}$ & 4.44$\times 10^{8}$ & 3.50$\times 10^{9}$ & ---                  & 0.23 & 5.24$\times 10^{47}$ \\
15& Yes & 1.15 & 0.83 & 0.63 & 1.85$\times 10^{-2}$ & 7.54$\times 10^{8}$ & 3.89$\times 10^{9}$ & 6.62$\times 10^{-3}$ & 0.20 & 2.63$\times 10^{48}$ \\
16& Yes & 1.26 & 0.54 & 0.67 & 6.81$\times 10^{-2}$ & 1.07$\times 10^{9}$ & 4.30$\times 10^{9}$ & 5.02$\times 10^{-3}$ & 0.19 & 2.36$\times 10^{49}$ \\
17&  No & 0.91 & 0.41 & 0.47 & 1.41$\times 10^{-2}$ & 4.36$\times 10^{8}$ & 1.16$\times 10^{9}$ & 5.44$\times 10^{-3}$ & 0.38 & 1.88$\times 10^{41}$ \\
18&  No & 1.38 & ---  & 0.21 & 5.76$\times 10^{-3}$ & 3.49$\times 10^{8}$ & 2.54$\times 10^{9}$ & ---                  & 0.25 & 7.95$\times 10^{46}$ \\ 
19&  No & 1.16 & 0.63 & 0.62 & 1.71$\times 10^{-2}$ & 7.02$\times 10^{8}$ & 2.08$\times 10^{9}$ & 4.85$\times 10^{-3}$ & 0.37 & 8.17$\times 10^{45}$ \\ 
20& Yes & 1.31 & 0.44 & 0.66 & 2.70$\times 10^{-2}$ & 1.25$\times 10^{9}$ & 2.80$\times 10^{9}$ & 3.65$\times 10^{-3}$ & 0.36 & 1.43$\times 10^{48}$ \\
21&  No & 0.92 & 0.42 & 0.46 & 1.39$\times 10^{-2}$ & 4.21$\times 10^{8}$ & 9.22$\times 10^{8}$ & 5.43$\times 10^{-3}$ & 0.49 & 1.90$\times 10^{39}$ \\
22&  No & 1.33 & ---  & 0.27 & 4.80$\times 10^{-3}$ & 2.48$\times 10^{8}$ & 1.32$\times 10^{9}$ & ---                  & 0.43 & 8.67$\times 10^{43}$ \\ 
23&  No & 1.17 & 0.50 & 0.62 & 1.73$\times 10^{-2}$ & 7.56$\times 10^{8}$ & 1.79$\times 10^{9}$ & 4.81$\times 10^{-3}$ & 0.34 & 9.95$\times 10^{45}$ \\
24& Yes & 1.30 & 0.44 & 0.67 & 2.62$\times 10^{-2}$ & 1.17$\times 10^{9}$ & 2.74$\times 10^{9}$ & 3.72$\times 10^{-3}$ & 0.28 & 9.38$\times 10^{47}$ \\
39& Yes & 0.42 & 0.23 & 0.16 & 2.43$\times 10^{-2}$ & 1.21$\times 10^{8}$ & 2.29$\times 10^{9}$ & 1.41$\times 10^{-2}$ & 0.75 & 2.04$\times 10^{48}$ \\
43& Yes & 0.41 & 0.22 & 0.16 & 2.26$\times 10^{-2}$ & 1.13$\times 10^{8}$ & 2.42$\times 10^{9}$ & 1.18$\times 10^{-2}$ & 0.49 & 2.17$\times 10^{48}$ \\ 
47&  No & 0.45 & 0.24 & 0.15 & 2.54$\times 10^{-3}$ & 1.09$\times 10^{8}$ & 6.83$\times 10^{8}$ & 1.63$\times 10^{-2}$ & 0.57 & 1.24$\times 10^{45}$ \\
48& Yes & 0.54 & ---  & 0.21 & 4.81$\times 10^{-2}$ & 8.81$\times 10^{7}$ & 2.68$\times 10^{9}$ & ---                  & 0.55 & 8.49$\times 10^{48}$ \\
51&  No & 0.44 & 0.17 & 0.16 & 5.63$\times 10^{-3}$ & 1.12$\times 10^{8}$ & 3.96$\times 10^{8}$ & 1.09$\times 10^{-2}$ & 0.65 & 2.75$\times 10^{43}$ \\
52& Yes & 0.67 & ---  & 0.13 & 4.96$\times 10^{-3}$ & 8.48$\times 10^{7}$ & 2.29$\times 10^{9}$ & ---                  & 0.64 & 8.12$\times 10^{47}$ \\
57& Yes & 0.85 & 0.29 & 0.32 & 3.13$\times 10^{-2}$ & 4.25$\times 10^{8}$ & 2.50$\times 10^{9}$ & 9.27$\times 10^{-3}$ & 0.38 & 6.09$\times 10^{48}$ \\
58& Yes & 1.20 & 0.27 & 0.32 & 7.80$\times 10^{-2}$ & 7.74$\times 10^{8}$ & 2.75$\times 10^{9}$ & 5.34$\times 10^{-3}$ & 0.41 & 2.03$\times 10^{49}$ \\
61& Yes & 0.85 & 0.28 & 0.32 & 2.52$\times 10^{-2}$ & 4.13$\times 10^{8}$ & 1.98$\times 10^{9}$ & 9.50$\times 10^{-3}$ & 0.33 & 4.70$\times 10^{48}$ \\
62& Yes & 1.21 & 0.15 & 0.30 & 9.15$\times 10^{-2}$ & 7.96$\times 10^{8}$ & 2.74$\times 10^{9}$ & 5.77$\times 10^{-3}$ & 0.45 & 2.53$\times 10^{49}$ \\
66& Yes & 1.20 & 0.15 & 0.35 & 5.15$\times 10^{-2}$ & 7.77$\times 10^{8}$ & 2.46$\times 10^{9}$ & 3.35$\times 10^{-3}$ & 0.74 & 1.86$\times 10^{49}$ \\
\noalign{\smallskip}
\hline
\hline
\end{tabular}
\end{center}
\end{table*}

\noindent  where $q=M_2/M_1$  is  the mass  ratio  of the  interacting
stars,  and  $a$  is the  binary  separation,  which  in our  case  is
$a=r_{\rm  min}$.  Nevertheless, we  emphasize that  Eq.~(\ref{RL}) is
only  valid for  circular orbits,  while  we are  dealing with  highly
eccentric  orbits.  Most  importantly,  this  expression  was  derived
assuming that the  interacting stars are spherical, which  in our case
is  not   true,  as  tidal  deformations  are   important  at  closest
approach. Thus,  instead of  directly using the  value of  $R_{\rm L}$
obtained from the expression above, we multiply it by a factor $\eta$,
which  takes into  account all  the non-modelled  effects.   Hence, we
expect  that the limiting  case separating  both dynamical  regimes is
$R_2 = \eta R_{\rm L}$, with $R_{\rm L}$ given by~Eq.~(\ref{RL}).

As  can be seen  in Fig.~\ref{plane},  the simple  argument previously
explained  works well when  $\eta \sim  0.95$ is  adopted, as  all the
interactions in which an eccentric  binary system is formed lay in the
lightest shaded area, whilst the  region of lateral collisions is also
nicely reproduced.  It is worth noting  as well that this value is the
same for  the simulations in which  a $0.8\, M_{\sun}$  white dwarf is
involved (left panel of Fig.~\ref{plane})  and those in which a $0.4\,
M_{\sun}$ star is adopted (right panel in the same figure). Hence, the
value of $\eta$ is robust, since it does not depend on the composition
of the  intervening white dwarf  (the $0.4\, M_{\sun}$ white  dwarf is
made  of  helium,  while  the  $0.8\,  M_{\sun}$  star  is  a  regular
carbon-oxygen  white  dwarf),  or  on  the  specific  details  of  the
interaction.  Note also  that the change in the  slope of the boundary
between both regions occurs at $\mu=0.4\, M_{\sun}$ for the left panel
and at  $\mu=0.2\, M_{\sun}$ for the right  panel of Fig.~\ref{plane}.
These  values correspond to  systems in  which both  intervening white
dwarfs  have equal  masses.  In  particular, if  we consider  the left
panel of  Fig.~\ref{plane}, stars  less massive than  $0.8\, M_{\sun}$
have larger  radii than  the $0.8\, M_{\sun}$  white dwarf due  to the
mass-radius  relation. Thus,  the  distances at  closest approach  for
which a Roche-lobe overflow episode occurs are larger, and the reverse
is  also true.   However, for  white dwarfs  more massive  than $0.8\,
M_{\sun}$ the  gravitational well is deeper,  and consequently lateral
collisions  occur  for  larger  periastron distances.   The  interplay
between these two effects determines the turn-off in this plane.

In Fig.~\ref{plane} we  also show the regions in  the plane defined by
the  reduced  mass and  periastron  distance  for  which the  physical
conditions for a  detonation are met during the  interaction. As shown
in  Figs.~\ref{CO-disrupted} and \ref{ONe-disrupted}  these conditions
always  occur in  the shocked  region  resulting from  the very  rapid
accretion phase  of the  disrupted less massive  star onto  the almost
rigid surface of the more massive white dwarf.  The regions in which a
detonation forms in the contact region between the surface of the more
massive star and the material accreted from the disrupted less massive
star are  located to the left  of the dashed  line.  The dotted-dashed
line  indicates the  edge  of the  region  for which  either the  less
massive star or both are  totally disrupted. We note, however, that if
the  more  massive white  dwarf  is  an  oxygen-neon white  dwarf  the
material of  the less  massive white dwarf  is not accreted  onto this
star, but it  bounces back. On the contrary if  both stars are massive
carbon-oxygen white  dwarfs, the system  is totally disrupted  and the
result  of the  interaction is  indeed a  super-Chandrasekhar  Type Ia
supernova.

\section{Physical properties of the interactions}
\label{remnants}

Quite  generally speaking,  and  except for  the  few cases  discussed
previously in which the material  of the less massive star is ejected,
or in those  cases in which both white  dwarfs are entirely disrupted,
in all  those simulations in which  a merger occurs,  the less massive
white dwarf  is accreted by  the massive companion, and  the resulting
configuration consists of a central compact object surrounded by a hot
corona, and a region where the debris of the interaction can be found.
The properties of the debris region  depend very much on the masses of
the interacting  white dwarfs.  In  particular, if a  direct collision
occurs,  the  final  configuration  is almost  spherically  symmetric,
whereas in  lateral collisions a  thick, heavy, rotationally-supported
disk is formed in all cases, but in those in which the two interacting
white dwarfs have  equal masses.  All these findings  are in agreement
with  those previously obtained  by \cite{PaEn10}.   We found  that in
these simulations  little mass is  ejected from the system,  except in
some direct collisions,  namely those with rather large  values of the
impact   parameter   $\beta$   ---  see   Tables~\ref{orbits-CO}   and
\ref{hydro-nodisrupt}.   The hot corona  corresponds to  material that
has been compressed and heated  during the collision, and therefore is
substantially   enriched    in   heavy   elements.     The   resulting
nucleosynthetic pattern follows closely that found in \cite{PaEn10}.

Table~\ref{hydro-nodisrupt}  shows some important  physical quantities
of the  simulations in which a  merger occurs but the  material of the
less massive  white dwarf is not  ejected from the  remnant.  That is,
those simulations in which the  remnant of the interaction is a single
bound  object,  with   the  configuration  previously  described.   In
particular, in this table we specify the masses of the central remnant
($M_{\rm WD}$),  of the corona  ($M_{\rm corona}$), and of  the debris
region ($M_{\rm debris}$).  Also  listed are the ejected mass ($M_{\rm
ej}$), as well as the radii of the corona and of the debris region ---
$R_{\rm corona}$  and $R_{\rm debris}$.  We considered  that the newly
formed white  dwarf is  made of  all the material  which rotates  as a
rigid solid plus the hot  corona, which rotates faster.  This includes
both  the unperturbed  more  massive star  and  the accreted  material
resulting from  the disrupted less  massive white dwarf.   Finally, we
also list the maximum temperature of the central remnant at the end of
our simulations --- $T_{\rm max}$ --- the maximum temperature achieved
during the most violent phase of the merger --- $T_{\rm peak}$ --- and
the energy released by nuclear reactions, $E_{\rm nuc}$.

\begin{table*}
\caption{Hydrodynamical results for the  simulations in which at least
  the material of one of the  colliding stars does not remain bound to
  the remnant.}
\label{hydro-ejected}
\begin{center}
\begin{tabular}{ccccccccc}
\hline
\hline
\noalign{\smallskip}
 Run & Remnant & $M_{\rm WD}$ & $M_{\rm debris}$ & $M_{\rm ej}$ & $v_{\rm ej}$      & $T_{\rm max}$ & $T_{\rm peak}$ & $E_{\rm nuc}$  \\
     &         & ($M_{\sun}$) & ($M_{\sun}$)     & ($M_{\sun}$) & ($\rm km/ \rm s$) & (K)           & (K)            & (erg) \\
\noalign{\smallskip}
\hline 
\hline
\noalign{\smallskip}
 3&  No & ---  & ---  & 1.80 & ---                 & ---                 & 1.70$\times 10^{10}$ & 1.42$\times 10^{51}$ \\
 4& Yes & 1.20 & 0.37 & 0.43 & 6.06$\times 10^{3}$ & 7.89$\times 10^{8}$ & 9.00$\times 10^{9}$  & 1.35$\times 10^{50}$ \\
 7&  No & ---  & ---  & 1.80 & ---                 & ---                 & 1.70$\times 10^{10}$ & 1.72$\times 10^{51}$ \\
 8& Yes & 1.20 & 0.39 & 0.41 & 5.95$\times 10^{3}$ & 7.66$\times 10^{8}$ & 8.51$\times 10^{9}$  & 1.42$\times 10^{50}$ \\
37& Yes & 0.37 & 0.06 & 0.16 & 3.19$\times 10^{3}$ & 9.16$\times 10^{7}$ & 2.75$\times 10^{9}$  & 1.61$\times 10^{49}$ \\
38&  No & ---  & ---  & 0.80 & ---                 & ---                 & 4.02$\times 10^{9}$  & 7.45$\times 10^{50}$ \\
40&  No & ---  & ---  & 0.80 & ---                 & ---                 & 4.03$\times 10^{9}$  & 7.24$\times 10^{50}$ \\                                  
41& Yes & 0.78 & 0.02 & 0.41 & 1.22$\times 10^{4}$ & 2.76$\times 10^{8}$ & 3.57$\times 10^{9}$  & 3.67$\times 10^{50}$ \\
42& Yes & 1.20 & 0.01 & 0.39 & 1.30$\times 10^{4}$ & 7.08$\times 10^{8}$ & 3.97$\times 10^{9}$  & 3.76$\times 10^{50}$ \\
44&  No & ---  & ---  & 0.80 & ---                 & ---                 & 4.06$\times 10^{9}$  & 7.38$\times 10^{50}$ \\    
45& Yes & 0.77 & 0.02 & 0.41 & 1.22$\times 10^{4}$ & 3.05$\times 10^{8}$ & 3.66$\times 10^{9}$  & 3.66$\times 10^{50}$ \\
46& Yes & 1.20 & 0.01 & 0.39 & 1.32$\times 10^{4}$ & 7.14$\times 10^{8}$ & 4.18$\times 10^{9}$  & 3.93$\times 10^{50}$ \\
49& Yes & 0.79 & 0.01 & 0.40 & 1.21$\times 10^{4}$ & 2.63$\times 10^{8}$ & 3.57$\times 10^{9}$  & 3.46$\times 10^{50}$ \\
50& Yes & 1.20 & 0.01 & 0.39 & 1.29$\times 10^{4}$ & 6.89$\times 10^{8}$ & 3.58$\times 10^{9}$  & 3.67$\times 10^{50}$ \\
53& Yes & 0.80 & 0.01 & 0.40 & 1.15$\times 10^{4}$ & 2.03$\times 10^{8}$ & 3.54$\times 10^{9}$  & 3.16$\times 10^{50}$ \\
54& Yes & 1.20 & 0.01 & 0.40 & 1.27$\times 10^{4}$ & 6.99$\times 10^{8}$ & 3.62$\times 10^{9}$  & 3.58$\times 10^{50}$ \\
\noalign{\smallskip}
\hline
\hline
\end{tabular}
\end{center}
\end{table*}

As can be seen in  Table~\ref{hydro-nodisrupt}, when the masses of the
intervening  white dwarfs are  rather different  --- namely,  when the
mass difference  between both stars is  $\ga 0.2 \,  M_{\sun}$ --- the
interaction  is more  violent.  This  is  the case,  for instance,  of
simulation number 5, in which two carbon-oxygen white dwarfs of masses
$0.8\,  M_{\sun}$ and  $0.6\, M_{\sun}$,  respectively,  are involved.
This occurs  because in  these cases the  less massive white  dwarf is
destroyed very rapidly in the  deep potential well of the more massive
one.  Moreover, in these simulations  the more massive white dwarf has
a  small  radius, which  leads  to  significant  accelerations of  the
material of  the disrupted less massive white  dwarf, and consequently
to stronger  interactions.  Furthermore, when the masses  of the white
dwarfs are  very different the  resulting white dwarf accretes  only a
small  percentage  of  the   less  massive  star.   If  the  dynamical
interaction results in a lateral  collision, a large amount of mass of
the disrupted less  massive white dwarf is incorporated  to the debris
region, whilst  if the interaction  results in a direct  collision the
mass  ejected   from  the  system  is  larger.    All  these  physical
considerations  are also  valid for  the simulations  in which  a very
massive oxygen-neon white  dwarf of mass $1.2 \,  M_{\sun}$ and a $0.8
\,  M_{\sun}$ carbon-oxygen  white dwarf  are involved  but  a lateral
collision occurs --- namely, simulations 12, 16, 20, and 24.  In these
simulations we find  that a larger fraction of  the less massive white
dwarf is ejected from the system,  while the mass of the debris region
is  also  significantly  larger.   On  the  contrary,  when  the  mass
difference  between  the  interacting  white  dwarfs  is  smaller  our
calculations show the interaction is  more gentle and a sizable amount
of  mass is  accreted on  the  undisrupted more  massive white  dwarf.
Finally, it  is important to realize  as well that  in all simulations
with small  impact parameters,  which produce lateral  collisions, the
number of  mass transfer episodes  is larger for decreasing  values in
the difference of masses of the interacting white dwarfs.

There is one simulation in which the remnant white dwarf is very close
to the  Chandraskhar mass, namely  simulation 18.  In  this simulation
two $0.8\, M_{\sun}$ white  dwarfs interact. Nevertheless, the remnant
is  rotating  rapidly (a  consequence  of  the  conversion of  orbital
angular momentum in rotational velocity  of the remnant), and thus the
central density is not extraordinarily large.  This prevents the onset
of  electron  captures on  $^{16}$O  in  the  densest regions  of  the
remnant.   Additionally,  in  some  simulations the  temperatures  and
densities attained during  the interaction are high enough  to drive a
detonation.   However, in  these simulations  the regions  in  which a
detonation is likely  to develop comprise a small  number of particles
and degeneracy  is rapidly lifted.   Consequently, in these  cases the
result of  the dynamical interaction  is not a  powerful thermonuclear
explosion, leading to a supernova.

There are a few simulations in which a corona is not formed. These are
simulations 2,  6, 10, 14,  18, 22, 48  and 52. All  these simulations
correspond to  interactions in which  two equal-mass white  dwarfs are
involved  (with  either  carbon-oxygen  or  helium  internal  chemical
compositions).   In these cases  both white  dwarfs are  disrupted and
merge forming  a new, more  massive, white dwarf.  In  particular, the
final remnant  is always  $\sim 35\%$ more  massive than  the original
white dwarf.   Moreover, in these simulations mixing  of the disrupted
white dwarfs  is extensive, and  the final temperature profile  of the
remnant is completely different from  that obtained in the rest of the
simulations.  In fact,  the final remnant is isothermal  and very hot.
Since the nuclear energy release in these simulations is modest, these
very high temperatures are the  result of compression of the disrupted
stars.

\begin{figure*}
\vspace{21.0cm}
\includegraphics{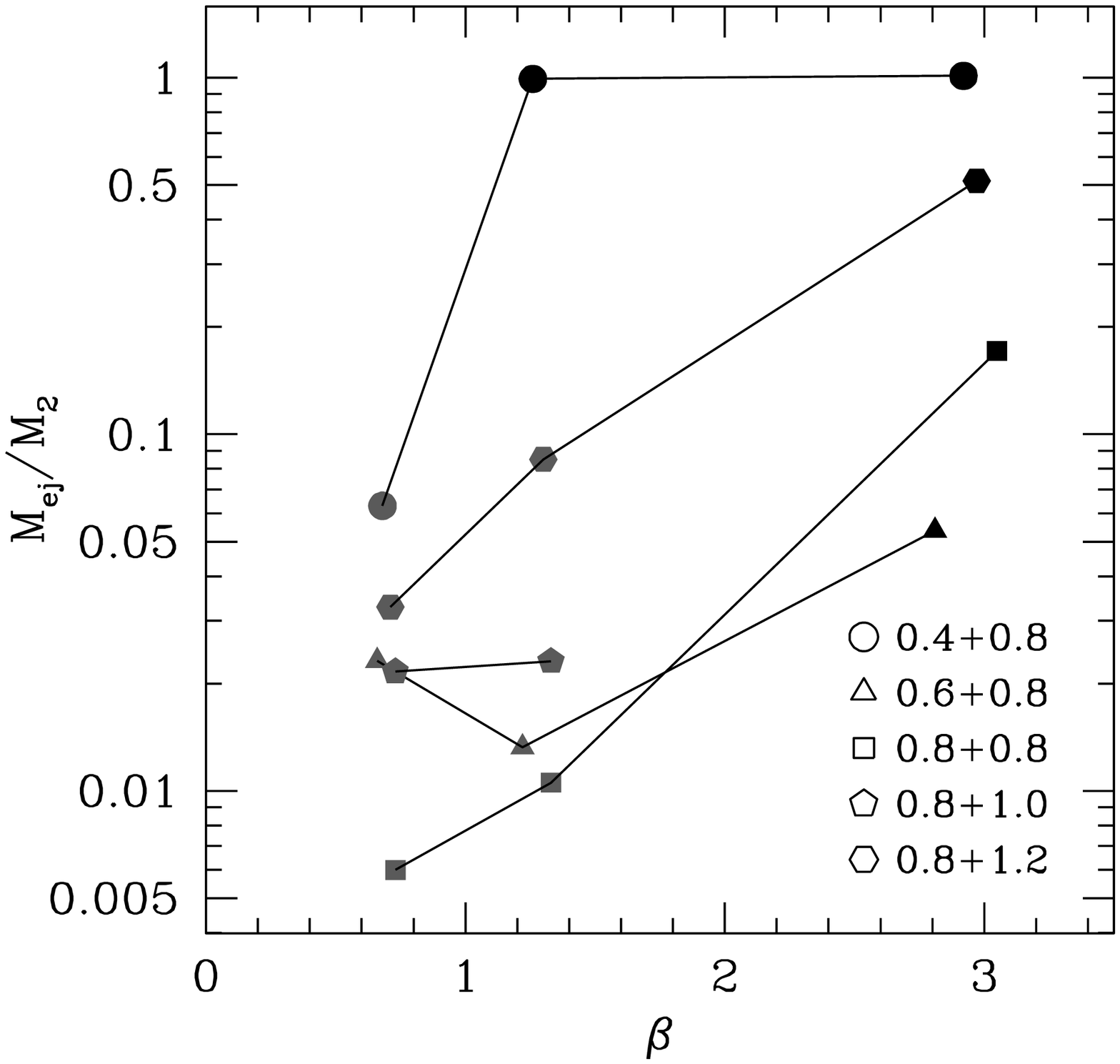}
\includegraphics{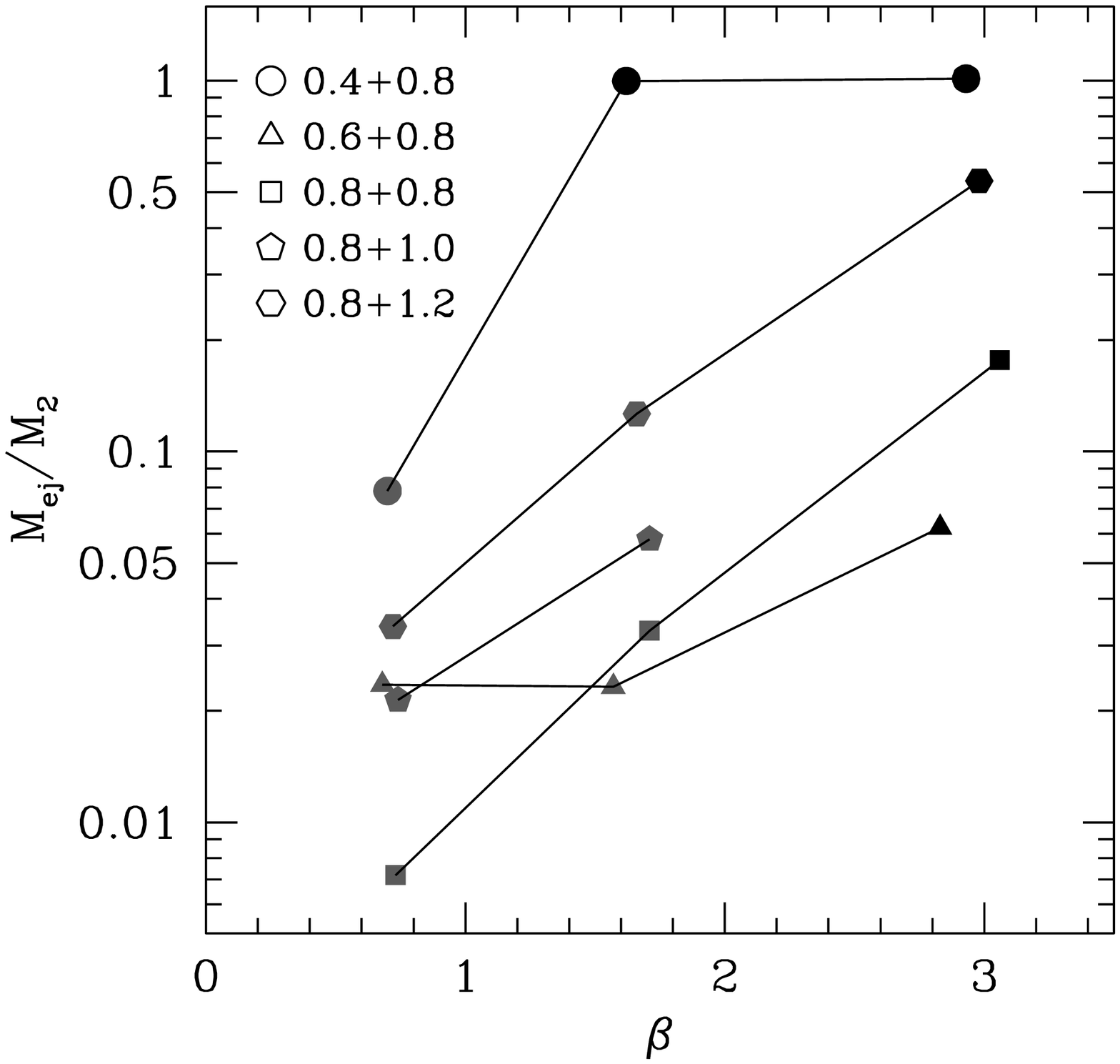}
\includegraphics{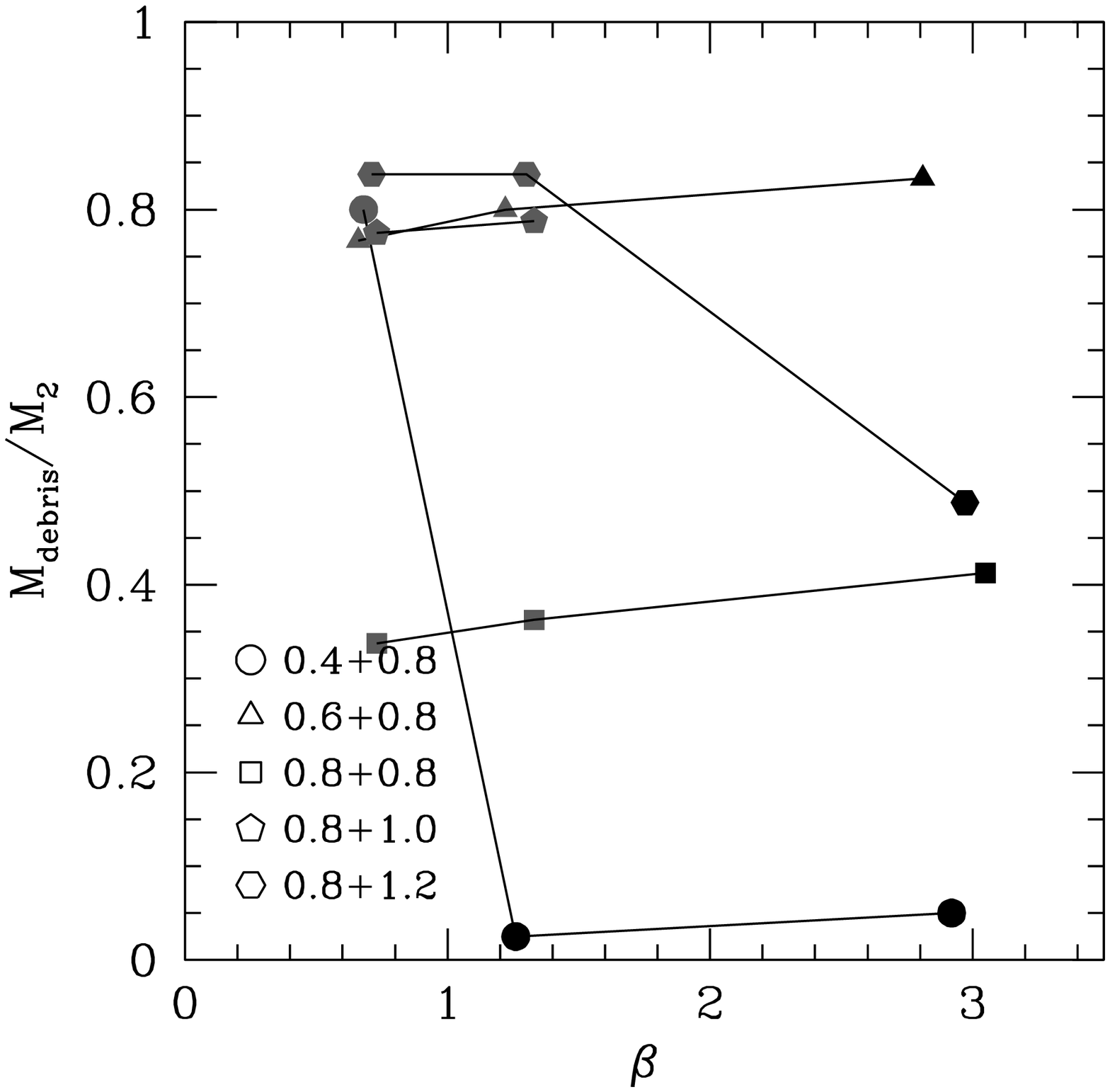}
\includegraphics{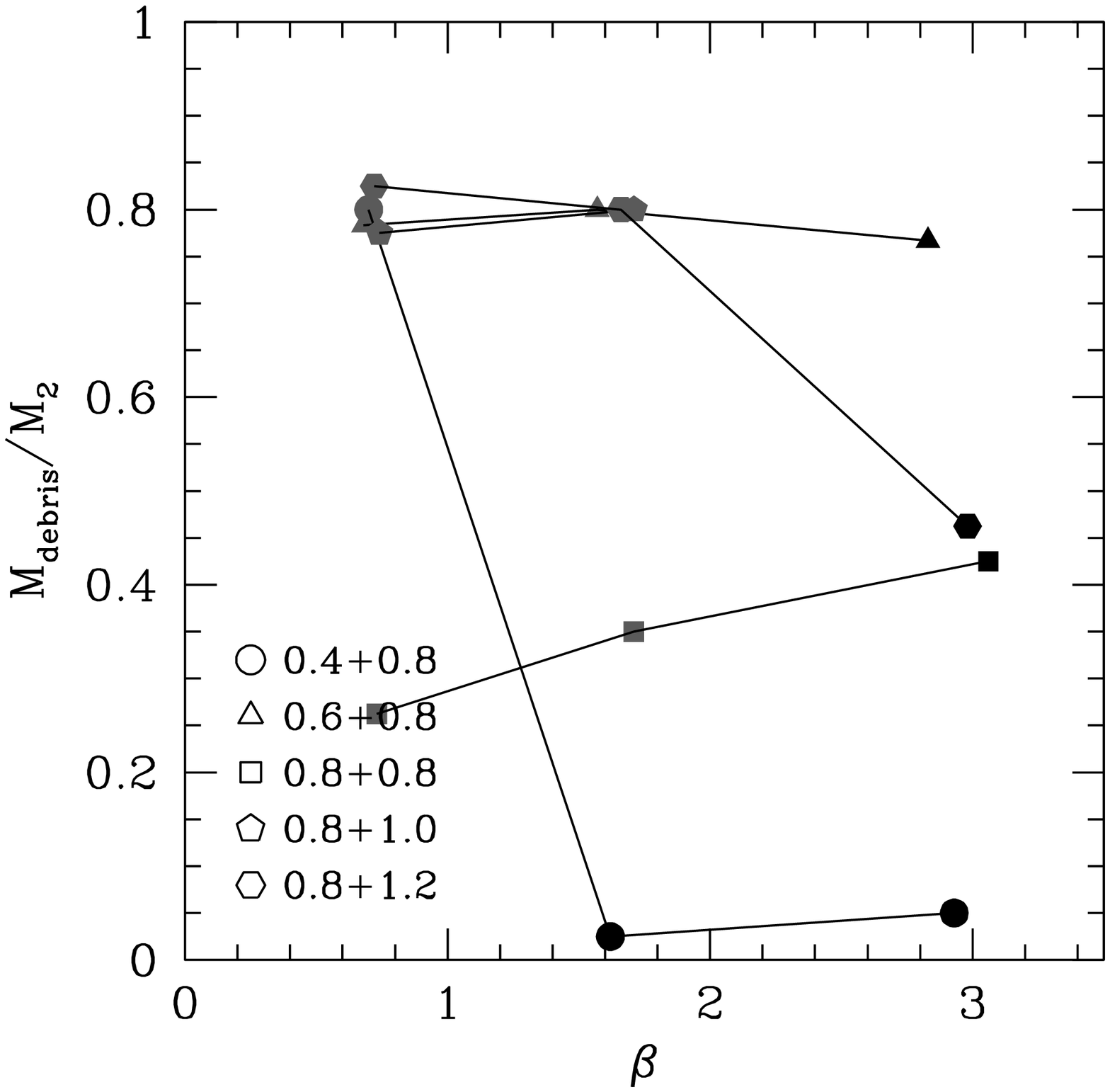}
\includegraphics{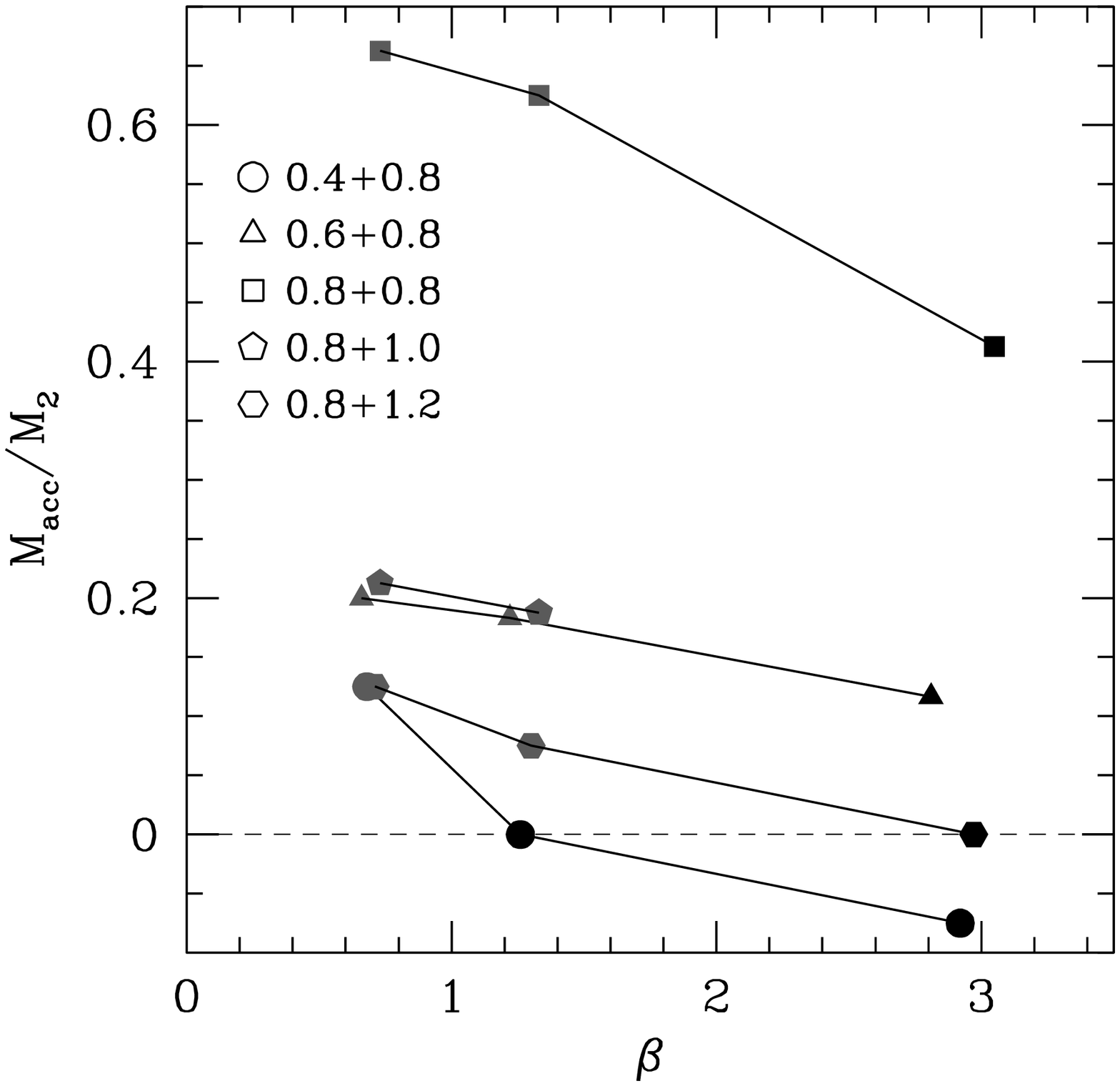}
\includegraphics{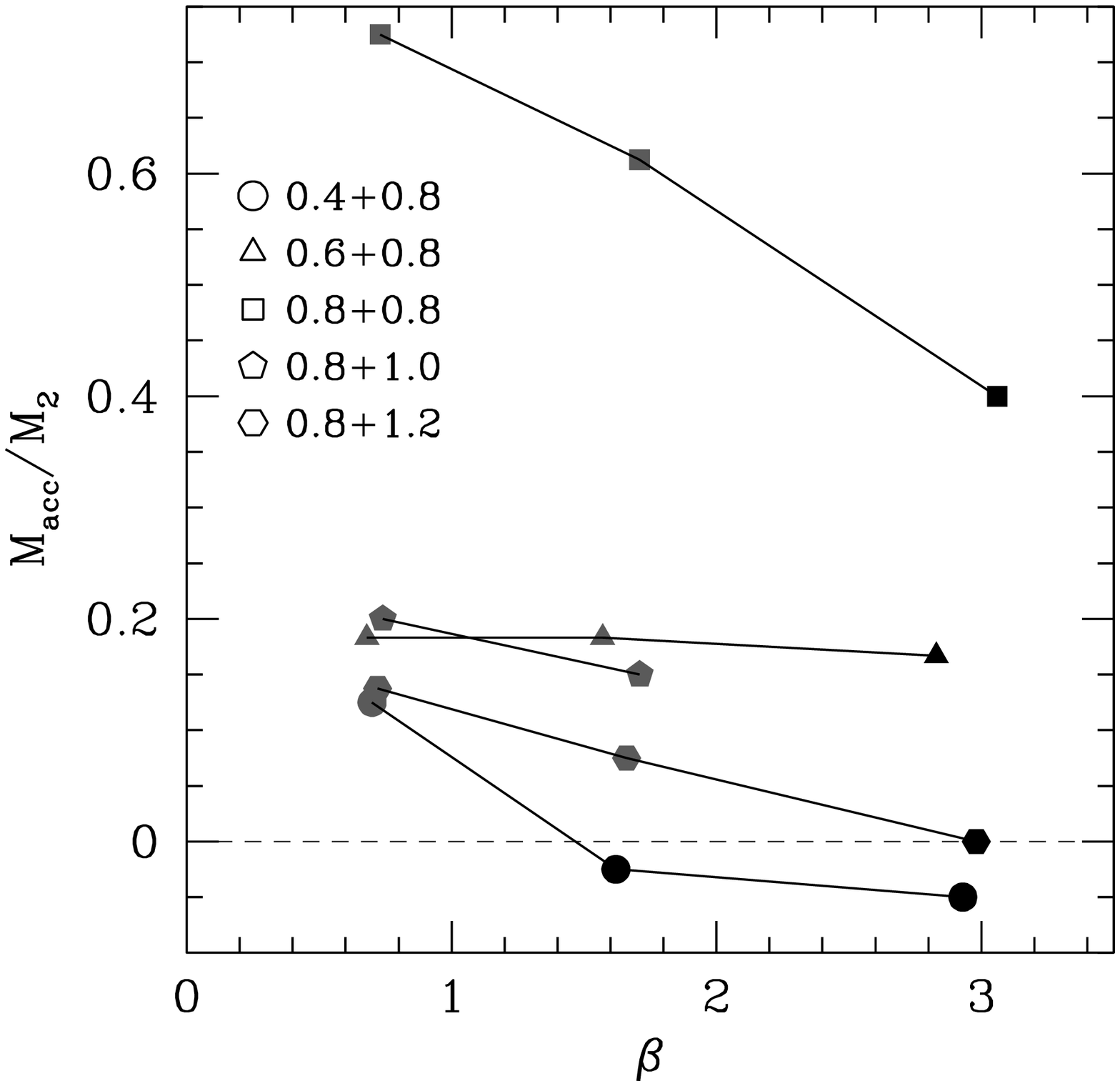}
\caption{Fraction of mass of the less massive white dwarf ejected (top
  panels), fraction of the disrupted star that forms the debris region
  (middle panels) and fraction of mass which is accreted onto the more
  massive  white dwarf  (bottom panels)  as a  function of  the impact
  parameter $\beta$,  for the simulations in which  a $0.8\, M_{\sun}$
  white  dwarf is  involved  and  a merger  occurs.   The left  panels
  correspond to the simulations in  which $\Delta y = 0.3 \, R_{\sun}$
  was adopted,  whereas in the  right panels the simulations  in which
  $\Delta y = 0.4 \, R_{\sun}$  was used are displayed.  The masses of
  the intervening white  dwarfs and the meaning of  the symbols can be
  found in the respective insets.  The grey symbols indicate a lateral
  collision,  while the black  ones indicate  a direct  collision. See
  text for additional details.}
\label{mejdeb1}
\end{figure*}

Table~\ref{hydro-ejected}  displays  the  results  obtained  in  those
simulations in which  either the material of the  less massive star is
not accreted  by the more  massive white dwarf,  and thus goes  to the
debris region or is ejected, or those calculations in which both stars
are totally disrupted.  Specifically, for  each run we first list if a
remnant  white dwarf  exists, the  corresponding mass  of  the remnant
white  dwarf, the mass  of the  debris region,  the ejected  mass, the
velocity  of the  ejecta, the  maximum  and peak  temperatures ---  as
previously defined --- attained  during the dynamical interaction, and
the nuclear energy released.

\begin{figure*}
\vspace{8.0cm}
\includegraphics{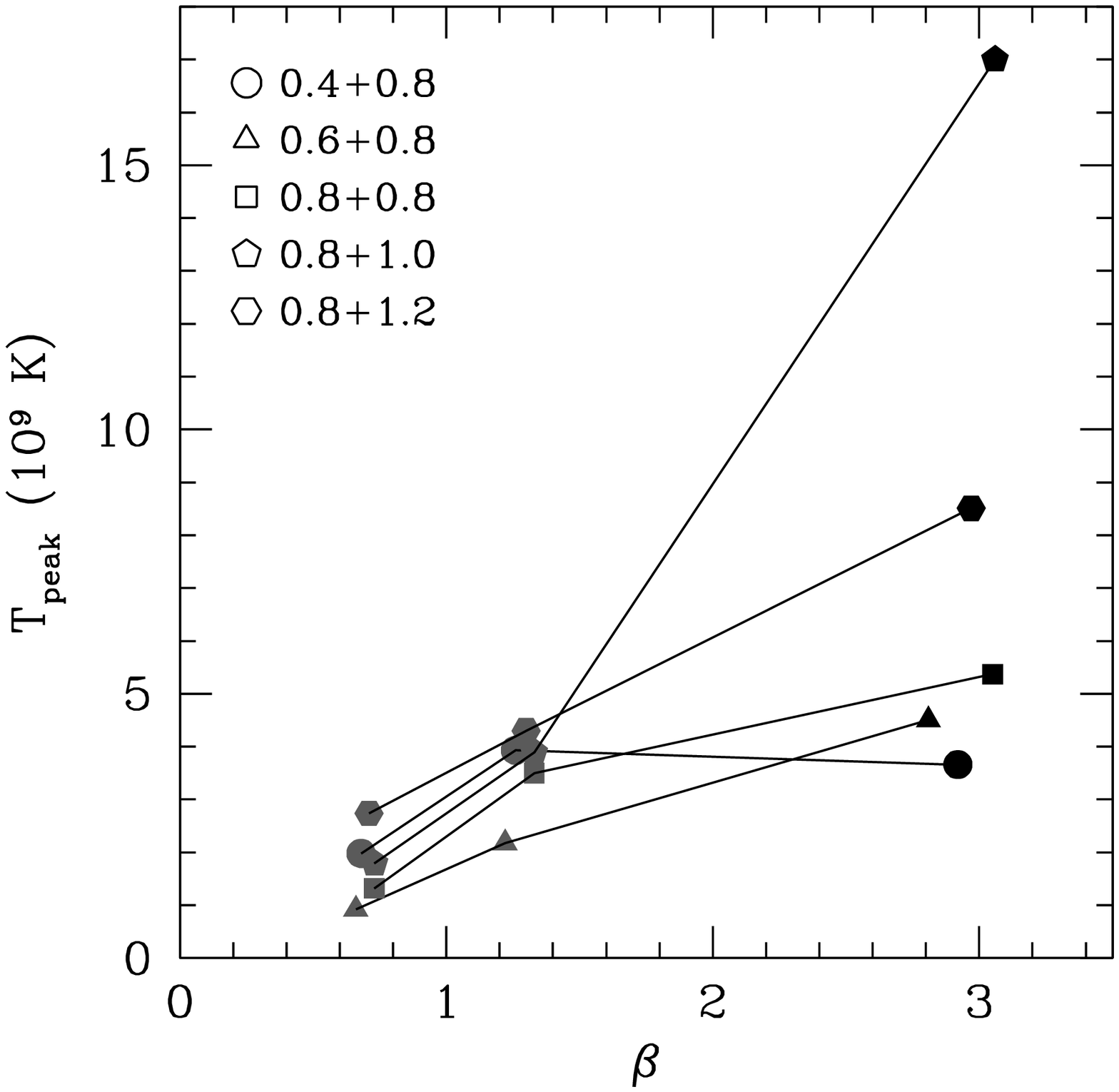}
\includegraphics{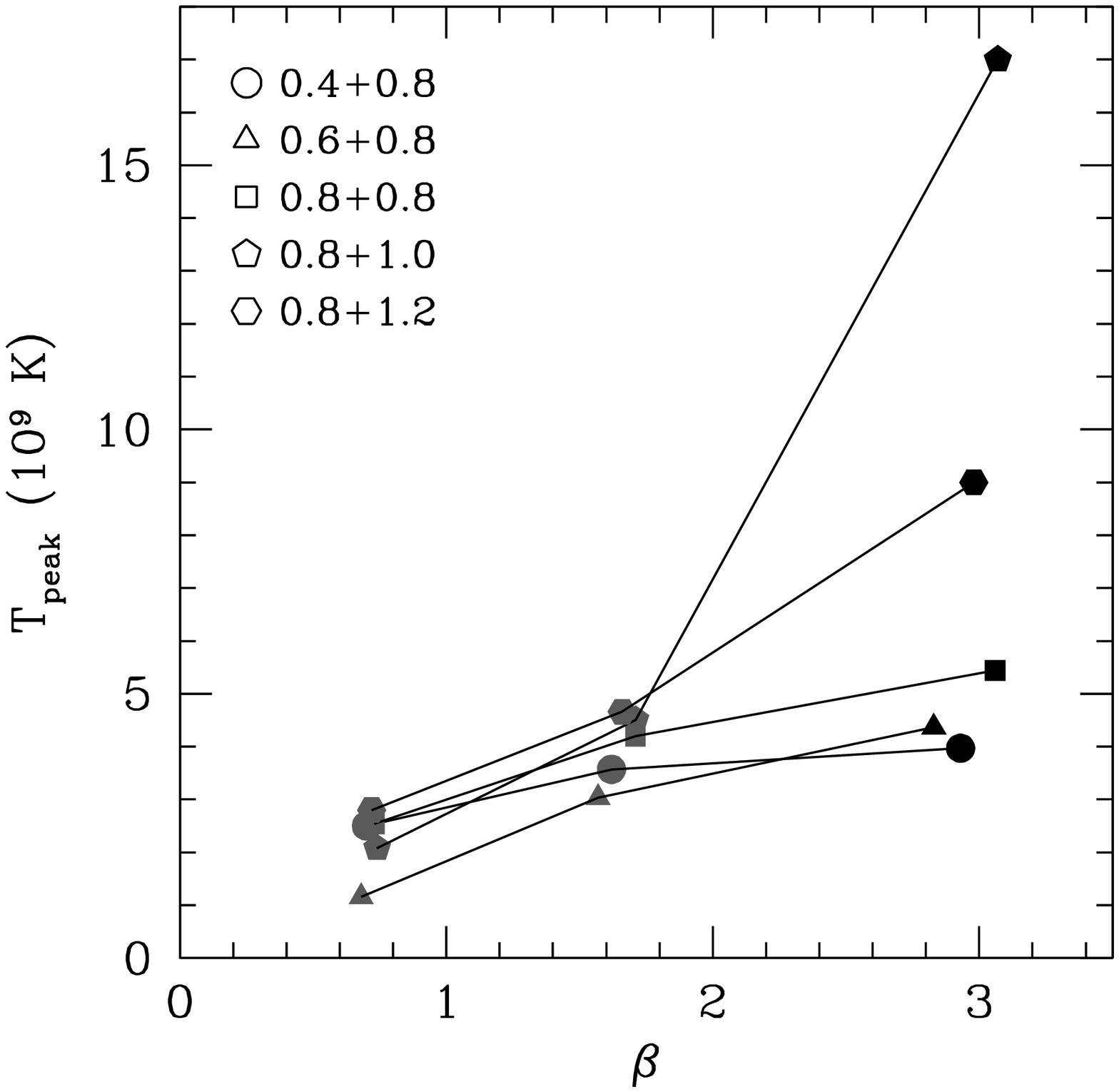}
\includegraphics{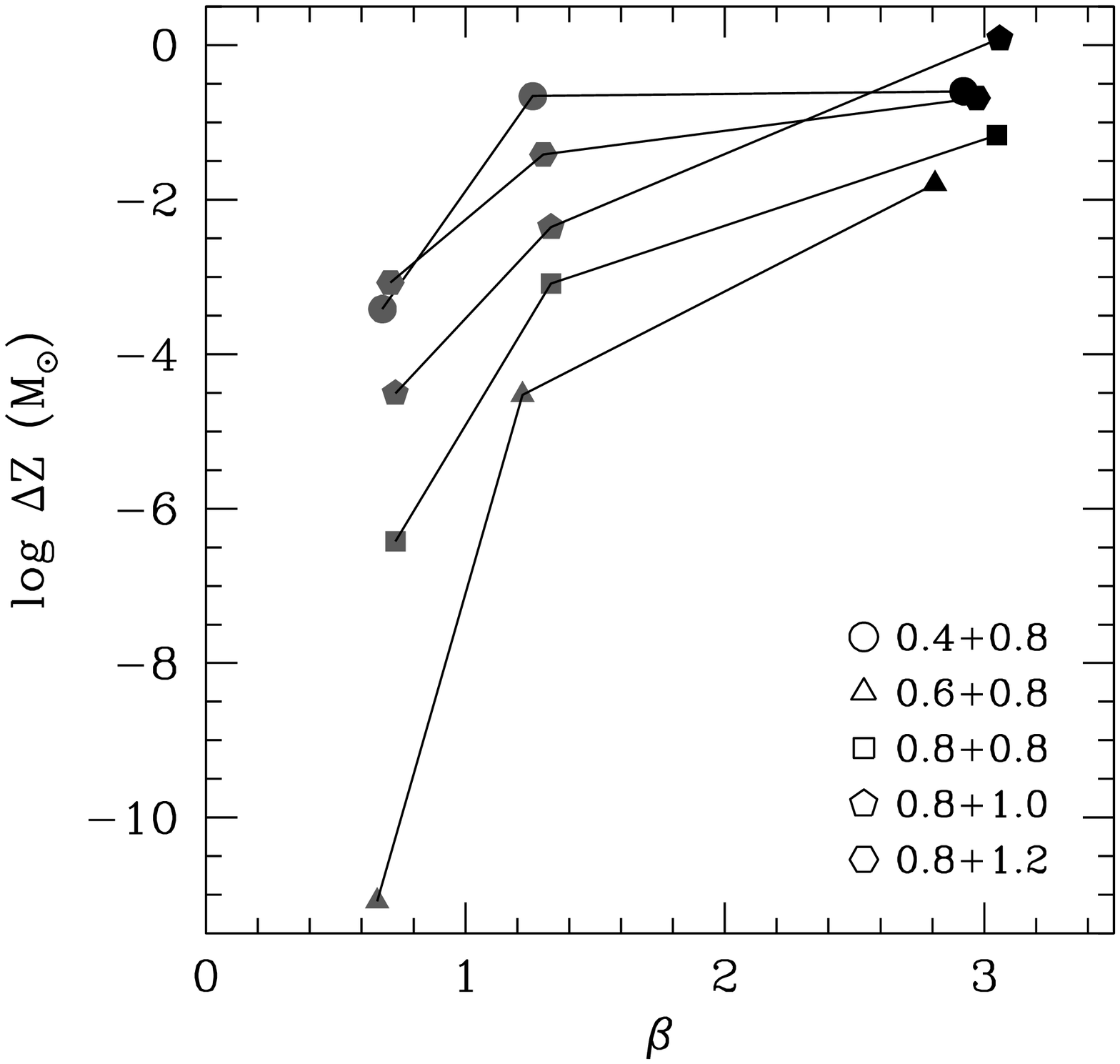}
\includegraphics{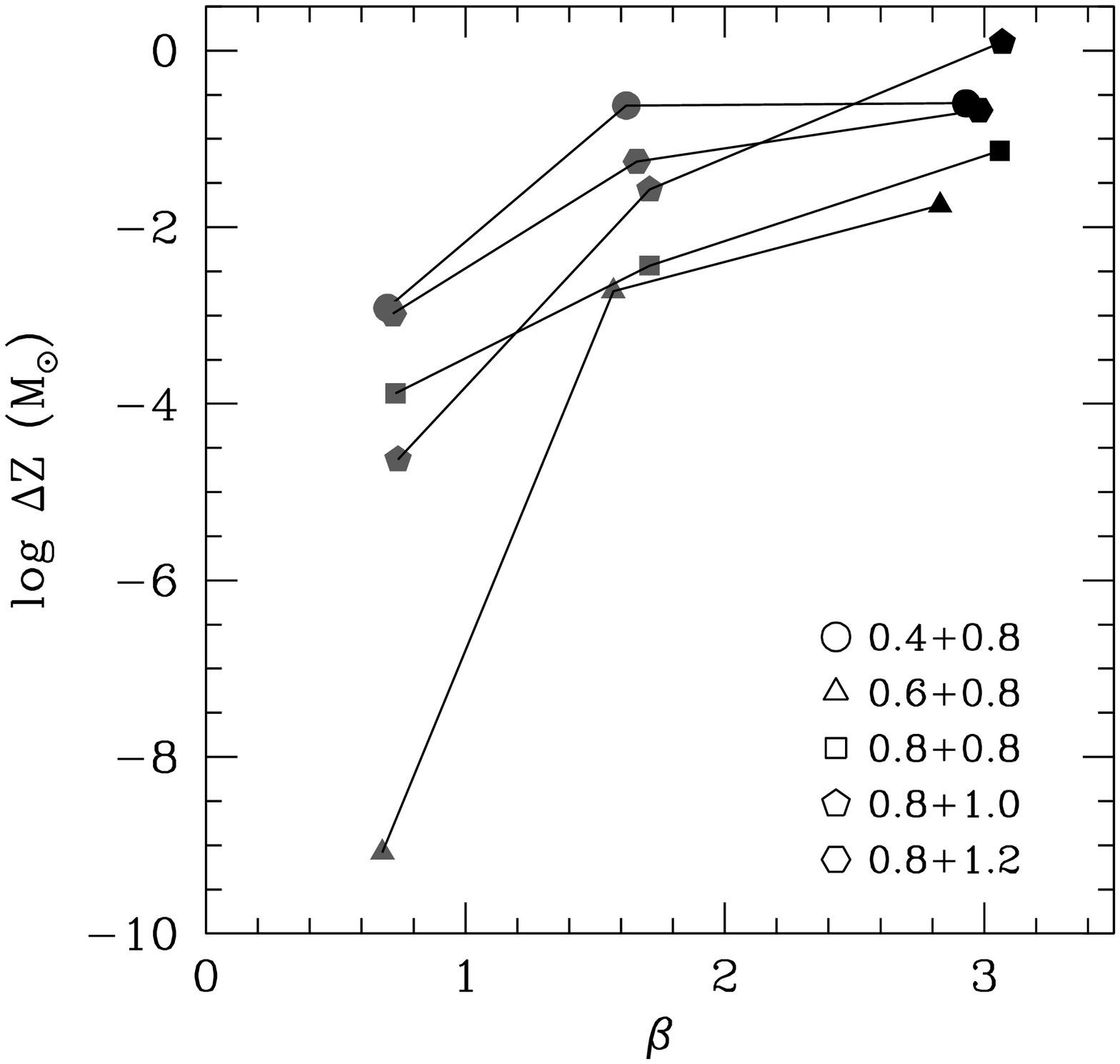}
\vspace{6.0cm}
\caption{Peak temperature achieved during the interaction (top panels)
  and  metallicity enhancement (bottom  panels) as  a function  of the
  impact  parameter $\beta$,  for the  simulations in  which  a $0.8\,
  M_{\sun}$ white  dwarf is  involved and a  merger occurs.   The left
  panels correspond  to the  simulations in which  $\Delta y =  0.3 \,
  R_{\sun}$ was  adopted, whereas in the right  panels the simulations
  in which $\Delta  y = 0.4 \, R_{\sun}$ was  used are displayed.  The
  masses of  the merging white dwarfs  and the meaning  of the symbols
  can be found in the  respective insets. As in Fig.~\ref{mejdeb1} the
  black symbols indicate  a direct collision, while the  grey ones are
  used to display lateral ones.}
\label{metaltemp1}
\end{figure*}

As can be seen, in most of these simulations a $0.4\, M_{\sun}$ helium
white dwarf  is involved, although there  are a few runs  in which the
intervening  star  is an  otherwise  regular,  not extremely  massive,
$0.8\,  M_{\sun}$ carbon-oxygen  white  dwarf.  This  is  the case  of
simulations  3, 4,  7 and  8.  This  can be  easily understood  as the
combination of two factors. On one hand, the material of the disrupted
less massive  helium white  dwarf has a  reduced Coulomb  barrier and,
hence,  nuclear reactions  are  more easily  driven  by the  dynamical
interaction.  On  the other, a  light-weight helium white dwarf  has a
reduced  gravity and,  consequently,  a larger  radius.   Thus, it  is
disrupted at larger  distances from the more massive  white dwarf, and
consequently the material flowing  to the primary can be significantly
compressed and heated during the  interaction.  All this results in an
explosive behavior.   However, in none of these  simulations the total
mass  of the  system is  larger  than Chandrasekhar's  mass, and  thus
although a powerful  nuclear outburst is powered by  the conversion of
gravitational energy  in thermal energy,  and a significant  amount of
nuclear energy is released (of the order of a few times $10^{50}$~erg)
in most cases this energy is invested in ejecting the shocked material
(approximately $0.4\, M_{\sun}$)  at significant velocities (typically
$10^4$~km/s).  Of these calculations there are two (namely, runs 3 and
7)  in which  the  outcome  is likely  a  super-Chandrasekhar Type  Ia
supernova,  since in this  case both  stars are  regular carbon-oxygen
white  dwarfs, although  rather massive  ($1.0\, M_{\sun}$  and $0.8\,
M_{\sun}$, respectively), the peak  temperatures are very large --- in
excess of $10^{10}$~K,  and no remnant is left  after the interaction.
In these two  simulations the released nuclear energies  are very high
($\sim  1.4\times  10^{51}$~erg   and  $\sim  1.7\times  10^{50}$~erg,
respectively),  but  these  values   should  be  considered  as  rough
estimates, as  nuclear statistical  equilibrium is not  implemented in
our  code.  In runs  4 and  8 a  heavy-weight oxygen-neon  white dwarf
tidally  disrupts a carbon-oxygen  white dwarf,  but the  more massive
star  remains bound,  and approximately  half of  the material  of the
disrupted star remains orbiting in the debris region, whereas the rest
of  the material is  ejected at  considerably large  velocities ($\sim
6\times 10^{3}$~km/s).  In these two cases the nuclear energy released
is somewhat smaller, of the order of $1.4\times 10^{50}$~erg.

All  this is  illustrated in  a different  way  in Fig.~\ref{mejdeb1},
where we show, for the  case in which a $0.8\, M_{\sun}$ carbon-oxygen
white  dwarf is involved,  the fraction  of mass  of the  less massive
white  dwarf ejected  during the  interaction ---  top panels  --- the
fraction of the disrupted star that forms the debris region --- middle
panels --- and  the mass fraction accreted by  the undisrupted massive
white  dwarf  ---  bottom panels  ---  as  a  function of  the  impact
parameter,  $\beta$. The  black, solid  symbols are  used to  depict a
direct collision,  while the grey  ones indicate a lateral  one.  Note
that direct collisions always occur  for large values of $\beta$.  The
left panels show the results obtained when $\Delta y =0.3\, R_{\sun}$,
while  the right panels  display the  results when  $\Delta y  = 0.4\,
R_{\sun}$.

\begin{table*}
\caption{Mass abundances  of selected  chemical elements (Mg,  Si, and
  Fe), for  the simulations in which  a $0.8\, M_{\sun}$  and a $0.6\,
  M_{\sun}$ white dwarfs collide and form a bound remnant.}
\begin{center}
\begin{tabular}{l r@{.}l r@{.}l r@{.}l r@{.}l cc}
\hline 
\hline 
\noalign{\smallskip}
 Run            & \multicolumn{2}{c}{1}                    & \multicolumn{2}{c}{5}                   & \multicolumn{2}{c}{9}                   & \multicolumn{2}{c}{13}                  & 17                  & 21                  \\
 $\beta$        & \multicolumn{2}{c}{2.83}                 & \multicolumn{2}{c}{2.81}                & \multicolumn{2}{c}{1.57}                & \multicolumn{2}{c}{1.22}                & 0.68                & 0.66                \\
 $T_{\rm peak}$ & \multicolumn {2}{c}{$4.37\times 10^{9}$} & \multicolumn{2}{c}{$4.50\times 10^{9}$} & \multicolumn{2}{c}{$3.03\times 10^{9}$} & \multicolumn{2}{c}{$2.18\times 10^{9}$} & $1.16\times 10^{9}$ & $9.22\times 10^{8}$ \\
\noalign{\smallskip}
\hline 
\hline
\multicolumn{11}{c}{Maximum abundances}\\
\hline
Mg                   & $2$ & $36\times 10^{-1}$  & $2$ & $33\times 10^{-1}$  & $2$ & $33\times 10^{-1}$  & $2$ & $12\times 10^{-1}$  & 8.00$\times 10^{-11}$ & 1.95$\times 10^{-14}$ \\
Si                   & $5$ & $86\times 10^{-1}$  & $5$ & $76\times 10^{-1}$  & $2$ & $39\times 10^{-1}$  & $1$ & $43\times 10^{-1}$  & 1.31$\times 10^{-16}$ & 1.51$\times 10^{-21}$ \\
Fe                   & $3$ & $54\times 10^{-3}$  & $1$ & $06\times 10^{-3}$  & $5$ & $03\times 10^{-24}$ & $6$ & $35\times 10^{-27}$ & 5.20$\times 10^{-34}$ & 5.20$\times 10^{-34}$ \\
\hline 
\multicolumn{11}{c}{Hot corona}\\
\hline
$\langle$Mg$\rangle$ & $1$ & $98\times 10^{-7}$  & $4$ & $14\times 10^{-7}$  & $9$ & $87\times 10^{-8}$  & $1$ & $53\times 10^{-12}$ & 1.50$\times 10^{-22}$ & 3.12$\times 10^{-24}$ \\
$\langle$Si$\rangle$ & $1$ & $50\times 10^{-9}$  & $6$ & $04\times 10^{-9}$  & $5$ & $51\times 10^{-10}$ & $6$ & $64\times 10^{-17}$ & 3.38$\times 10^{-31}$ & 1.03$\times 10^{-33}$ \\
$\langle$Fe$\rangle$ & $5$ & $20\times 10^{-34}$ & $5$ & $20\times 10^{-34}$ & $5$ & $19\times 10^{-34}$ & $5$ & $20\times 10^{-34}$ & 5.15$\times 10^{-34}$ & 5.09$\times 10^{-34}$ \\
\hline 
\multicolumn{11}{c}{Debris region}\\
\hline
$\langle$Mg$\rangle$ & $2$ & $97\times 10^{-3}$  & $2$ & $74\times 10^{-3}$  & $1$ & $16\times 10^{-3}$  & $2$ & $48\times 10^{-5}$  & 1.04$\times 10^{-14}$ & 2.28$\times 10^{-18}$ \\
$\langle$Si$\rangle$ & $7$ & $17\times 10^{-3}$  & $5$ & $71\times 10^{-3}$  & $5$ & $84\times 10^{-4}$  & $1$ & $16\times 10^{-5}$  & 1.57$\times 10^{-20}$ & 1.30$\times 10^{-25}$ \\
$\langle$Fe$\rangle$ & $5$ & $14\times 10^{-7}$  & $2$ & $60\times 10^{-7}$  & $4$ & $49\times 10^{-28}$ & $3$ & $81\times 10^{-31}$ & 4.63$\times 10^{-34}$ & 4.42$\times 10^{-34}$ \\
\hline
\multicolumn{11}{c}{Ejecta}\\
\hline
$\langle$Mg$\rangle$ & $5$ & $92\times 10^{-2}$  & $5$ & $81\times 10^{-2}$  & $1$ & $58\times 10^{-2}$  & $5$ & $72\times 10^{-10}$ & 6.38$\times 10^{-18}$ & 3.24$\times 10^{-17}$ \\
$\langle$Si$\rangle$ & $1$ & $27\times 10^{-1}$  & $1$ & $29\times 10^{-1}$  & $9$ & $27\times 10^{-3}$  & $3$ & $84\times 10^{-14}$ & 3.07$\times 10^{-25}$ & 2.73$\times 10^{-24}$ \\
$\langle$Fe$\rangle$ & $2$ & $16\times 10^{-6}$  & $3$ & $15\times 10^{-6}$  & $2$ & $97\times 10^{-27}$ & $4$ & $10\times 10^{-34}$ & 5.20$\times 10^{-34}$ & 5.20$\times 10^{-34}$ \\
\noalign{\smallskip}
\hline
\hline
\end{tabular}
\end{center}
\label{nucleo}
\end{table*}

We start  discussing the top  panels of Fig.~\ref{mejdeb1}. As  can be
seen  in  these  panels,  the  fraction of  mass  ejected  during  the
interaction is small when a  lateral collision occurs, of the order of
10\% at most.   Moreover, for a given set of  masses, the mass ejected
during the interaction increases for increasing values of $\beta$. The
only exception is the case in  which two white dwarfs of masses $0.8\,
M_{\sun}$ and  $0.6\, M_{\sun}$ interact.  For these  runs the ejected
mass  first   decreases  slightly  as  $\beta$   increases,  and  then
increases,  as in  the rest  of the  simulations. It  is  important to
realize that the  mass ejected from the system  depends noticeably not
only on  the masses of the  interacting white dwarfs, but  also on the
peak temperatures  achieved during the first and  most violent moments
of the  interaction ---  which are larger  when both stars  are rather
massive and  have similar masses  --- as well  as on other  details of
each  simulation, like  the  chemical composition  of the  interacting
white dwarfs.   This explains why we  obtain in this  case a different
behavior.  Note as well that  the mass ejected in direct collisions is
considerably larger,  and in some  extreme cases --- for  instance, in
the  case in which  a $0.8\,  M_{\sun}$ and  a $0.4\,  M_{\sun}$ white
dwarf interact  --- all the mass  of the less massive  star is ejected
from the system. Additionally, comparing the top left and right panels
it  turns out  that  all this  is  nearly independent  of the  adopted
initial distance along  the $y$-axis, $\Delta y$.  This  can be easily
understood.  If $\beta$ is fixed, then all that $\Delta y$ controls is
the eccentricity  of the encounter,  and the eccentricity  varies very
little for the $\Delta y$ values chosen.

Now we turn our attention to  the mass of the debris region --- middle
panels of Fig.~\ref{mejdeb1}.  As can  be seen, when the mass contrast
is large  (say $\ga 0.3  \, M_{\sun}$) the  mass of the  debris region
decreases for increasing values of $\beta$, whereas in all other cases
the fraction  of the less massive  white dwarf which goes  to form the
debris  region slightly  increases as  for larger  values  of $\beta$.
Note as well that, as it occurs with the mass ejected from the system,
the  general trend  is  nearly  independent of  the  adopted value  of
$\Delta y$.  Finally, we discuss the mass accreted by the more massive
white dwarf  --- bottom  panels of Fig.~\ref{mejdeb1}.   As previously
mentioned, the  final mass of the  remnant white dwarf is  the mass of
the undisrupted  massive component of the  system and the  mass of the
hot  corona that is  formed during  the interaction.   Our simulations
demonstrate that for strong impacts --- or, equivalently, large values
of $\beta$  --- the accreted  mass decreases substantially  as $\beta$
increases. Actually, the accreted  mass can be negative for relatively
large values of $\beta$. This means that actually some material of the
more massive white dwarf is  removed during interaction and is ejected
or incorporated into  the debris region.  This occurs  for the case in
which a  $0.4\, M_{\sun}$ helium  white dwarf interacts with  a $0.8\,
M_{\sun}$ carbon-oxygen  one. In the rest  of the cases  --- except in
that  in  which  two  equal-mass  $0.8\, M_{\sun}$  interact  ---  the
accreted mass is $\sim 20\%$ in lateral collisions and much smaller in
direct ones.  As mentioned, the  exception to this general rule is the
previously mentioned  case in which two $0.8\,  M_{\sun}$ white dwarfs
collide. In this case the fraction of mass incorporated to the remnant
is considerably larger, typically  $\sim 50\%$.  Finally, we emphasize
that, again,  all these considerations are independent  of $\Delta y$,
for the same reason previously discussed.

Finally, to close this section we discuss Fig.~\ref{metaltemp1}, where
we show the peak temperatures reached during the interactions in which
a  $0.8\, M_{\sun}$  white dwarf  is involved,  as a  function  of the
impact  parameter $\beta$  --- top  panels ---  and  the corresponding
metallicity   enhancements  due  to   nuclear  reactions   ---  bottom
panels. As  in Fig.~\ref{mejdeb1}  the left panels  show the  cases in
which $\Delta y=0.3 \, R_{\sun}$ was adopted, whereas the right panels
show the cases  in which $\Delta y=0.4 \, R_{\sun}$  was used. We also
use the  same convention adopted  in Fig.~\ref{mejdeb1}, and  the grey
symbols indicate  a lateral collision,  while the black ones  denote a
direct one. As in Fig.~\ref{mejdeb1} the masses of the colliding white
dwarf are  indicated in the insets of  each figure. As can  be seen in
this figure, the peak  temperatures reached during the interaction ---
top panels --- increase for increasing values of $\beta$ in almost all
cases.  It  is worth  noticing that,  except for the  case in  which a
$0.8\, M_{\sun}$  and a $1.0  \, M_{\sun}$ white dwarfs  interact, the
relationship  between $T_{\rm  peak}$  and $\beta$  is almost  linear.
Note that in the case  in which two rather massive carbon-oxygen white
dwarfs  of  masses  $1.0\,  M_{\sun}$ and  $0.8\,  M_{\sun}$  directly
collide  the  resulting  peak  temperature is  rather  large,  $T_{\rm
peak}\simeq 1.7  \times 10^{10}$.  As previously found  for the masses
ejected  from  the  system,  accreted  to  the  central  remnant,  and
incorporated  to   the  debris  region,  the  behavior   of  the  peak
temperature  as a  function of  $\beta$ is  almost independent  of the
adopted value of $\Delta y$. Another important characteristic is that,
in general,  the larger the total  mass of the system,  the higher the
peak temperature reached during the interaction. Additionally, we find
that  the more  violent  the interaction  ---  and, consequently,  the
larger the  mass transfer  rate --- the  higher the  peak temperatures
are. This is a natural consequence of the matter of the disrupted less
massive star being more  rapidly compressed in these interactions.  We
now turn  our attention to the  bottom panels of this  figure, were we
show  the  metallicity enhancement,  $\Delta  Z$,  resulting from  our
simulations.   Here  we define  $\Delta  Z$ as  the  sum  of the  mass
abundances of all elements that were not present in the original white
dwarfs plus helium.  For example,  if the two intervening white dwarfs
are  made  of carbon-oxygen  $\Delta  Z$ stands  for  the  sum of  the
abundances of  all elements  heavier than oxygen.   As can be  seen in
these panels,  the metallicity enhancement increases  abruptly until a
critical value of $\beta$, while for values of $\beta$ larger than the
critical  one the metallicity  enhancement also  increases but  with a
shallower   slope.    Examining  the   left   and   right  panels   of
Fig.~\ref{metaltemp1}  it  turns out  that  this  critical value  lies
between $\beta\sim 1.2$  and $\sim 1.6$.  All this  is, obviously, the
result  of   the  high  temperatures  reached   during  the  dynamical
interactions,  and hence  the metallicity  enhancement shows  the same
overall behavior observed for  the peak temperatures.  However, due to
the  strong   dependence  of  the  thermonuclear   reaction  rates  on
temperature  the slopes  of the  relationship between  $\Delta  Z$ and
$\beta$   are  much   steeper  (note   that  the   bottom   panels  of
Fig.~\ref{metaltemp1} have a logarithmic scale).

It is  as well interesting  to note that the  nucleosynthetic endpoint
depends sensitively  on $\beta$,  as it should  be expected.   This is
illustrated in Table~\ref{nucleo} where we list the abundances of some
selected  chemical  species for  the  simulations  in  which a  $0.8\,
M_{\sun}$ and a $0.6\, M_{\sun}$ white dwarfs collide and form a bound
remnant, which  is a representative case of  the calculations reported
here.   Specifically, for  these runs  we list  the  respective impact
parameter, the peak temperature reached during the interaction and the
abundances (by mass) of magnesium, silicon and iron in the hot corona,
in the  debris region  and in  the ejecta.  We  also list  the maximum
abundances of  these elements  found in the  entire remnant.   We note
that these abundances  are in line with those  found by \cite{PaEn10}.
The first thing to be noted in this table is that the larger the value
of  $\beta$ ---  and, consequently,  of  $T_{\rm peak}$  --- the  more
extensive nuclear processing is  and the larger the maximum abundances
of  these elements.   Secondly, it  is important  to realize  that the
average abundances of these elements  are largest in the ejecta, while
they are  rather small in the  hot corona and have  sizeable values in
the  debris region.   This,  in turn,  is  a consequence  of the  very
intensive nuclear processing of the shocked material hitting the rigid
surface  of the  more massive  white  dwarf and  acquiring very  large
velocities  during the  interaction, which  results in  the subsequent
ejection of these particles. This  is particularly true for runs 1 and
5. Note that for these simulations the iron and silicon abundances are
very  large, and that  the silicon  abundance is  larger than  that of
magnesium and  iron, indicating that nuclear processing  has been very
extensive, but has not been  complete.  Magnesium is the most abundant
isotope in the  ejecta of run 9, and  its abundance decreases abruptly
for succesive runs, while silicon is the most abundant element maximum
in  the ejecta  of  runs 1  and  5.  For  these  simulations the  iron
abundance is also maximum,  but for $\beta\la 1.6$ becomes negligible.
The  same  behavior is  found  for  the  debris region.   Finally,  we
emphasize that  the iron  abundance is negligible  in the  hot corona,
independently  of  the value  of  $\beta$,  whilst  the magnesium  and
silicon abundances are modest there.

\begin{table}
\caption{$^{56}$Ni production in the  collisions that have undergone a
  detonation.}
\label{niquel}
\begin{center}
\begin{tabular}{l c r@{.}l r@{.}l}
\hline
\hline
\noalign{\smallskip}
 Run & Ejection &  \multicolumn{2}{c}{$M_{\rm Ni}$} & \multicolumn{2}{c}{$L$} \\
     &          &  \multicolumn{2}{c}{($M_{\sun}$)} & \multicolumn{2}{c}{erg/s} \\
\noalign{\smallskip}
\hline 
\hline
\noalign{\smallskip}
 1& No & 8 & 65$\times 10^{-8}$  & 1 & 79$\times 10^{36}$ \\
 2& No & 4 & 47$\times 10^{-3}$  & 9 & 23$\times 10^{40}$ \\
 3&  2 & 7 & 25$\times 10^{-1}$  & 1 & 50$\times 10^{43}$ \\
 4&  1 & 6 & 60$\times 10^{-2}$  & 1 & 36$\times 10^{42}$ \\
 5& No & 2 & 74$\times 10^{-8}$  & 5 & 66$\times 10^{35}$ \\  
 6& No & 3 & 67$\times 10^{-3}$  & 7 & 58$\times 10^{40}$ \\
 7&  2 & 7 & 15$\times 10^{-1}$  & 1 & 48$\times 10^{43}$ \\
 8&  1 & 6 & 32$\times 10^{-2}$  & 1 & 31$\times 10^{42}$ \\
 9& No & 7 & 89$\times 10^{-34}$ & 1 & 63$\times 10^{10}$ \\  
10& No & 4 & 52$\times 10^{-10}$ & 9 & 33$\times 10^{33}$ \\
11& No & 7 & 94$\times 10^{-7}$  & 1 & 64$\times 10^{37}$ \\
12& No & 1 & 09$\times 10^{-5}$  & 2 & 25$\times 10^{38}$ \\
14& No & 7 & 44$\times 10^{-16}$ & 1 & 54$\times 10^{28}$ \\
15& No & 1 & 56$\times 10^{-13}$ & 3 & 22$\times 10^{30}$ \\
16& No & 9 & 70$\times 10^{-10}$ & 2 & 00$\times 10^{34}$ \\
20& No & 1 & 12$\times 10^{-33}$ & 2 & 31$\times 10^{10}$ \\
24& No & 1 & 12$\times 10^{-33}$ & 2 & 31$\times 10^{10}$ \\
37&  1 & 1 & 31$\times 10^{-9}$  & 2 & 71$\times 10^{34}$ \\
38&  2 & 8 & 84$\times 10^{-4}$  & 1 & 83$\times 10^{40}$ \\
39& No & 5 & 57$\times 10^{-16}$ & 1 & 15$\times 10^{28}$ \\
40&  2 & 1 & 64$\times 10^{-3}$  & 3 & 39$\times 10^{40}$ \\
41&  1 & 8 & 04$\times 10^{-4}$  & 1 & 66$\times 10^{40}$ \\
42&  1 & 1 & 18$\times 10^{-3}$  & 2 & 44$\times 10^{40}$ \\
43& No & 4 & 97$\times 10^{-14}$ & 1 & 03$\times 10^{30}$ \\ 
44&  2 & 4 & 68$\times 10^{-3}$  & 9 & 67$\times 10^{40}$ \\
45&  1 & 7 & 60$\times 10^{-4}$  & 1 & 57$\times 10^{40}$ \\
46&  1 & 2 & 21$\times 10^{-3}$  & 4 & 56$\times 10^{40}$ \\
48& No & 2 & 76$\times 10^{-11}$ & 5 & 70$\times 10^{32}$ \\
49&  1 & 5 & 00$\times 10^{-4}$  & 1 & 03$\times 10^{40}$ \\
50& No & 9 & 36$\times 10^{-4}$  & 1 & 93$\times 10^{40}$ \\
52& No & 1 & 62$\times 10^{-17}$ & 3 & 35$\times 10^{26}$ \\
53&  1 & 2 & 00$\times 10^{-4}$  & 4 & 13$\times 10^{39}$ \\
54&  1 & 8 & 44$\times 10^{-4}$  & 1 & 74$\times 10^{40}$ \\
57& No & 6 & 98$\times 10^{-14}$ & 1 & 44$\times 10^{30}$ \\
58& No & 5 & 61$\times 10^{-10}$ & 1 & 16$\times 10^{34}$ \\
61& No & 2 & 37$\times 10^{-16}$ & 4 & 89$\times 10^{27}$ \\
62& No & 3 & 27$\times 10^{-9}$  & 6 & 75$\times 10^{34}$ \\
66& No & 7 & 30$\times 10^{-12}$ & 1 & 51$\times 10^{32}$ \\
\noalign{\smallskip}
\hline
\hline
\end{tabular}
\end{center}
\label{nickel}
\end{table}

Finally, table~\ref{nickel} lists the mass of $^{56}$Ni synthesized in
those simulations in which  a detonation occurs, and the corresponding
energy release according to  the simple scaling law of \cite{Arnett82}
--- see, for instance,  \cite{Branch92}. As can be seen  in this table
the amount of $^{56}$Ni synthesized in these simulations, and thus the
corresponding luminosities, is  small in almost all the  cases, with a
few exceptions, namely those  simulations in which either the material
of the lightest intervening white dwarf or of both stars is ejected as
a consequence  of the dynamical interaction.  In  particular, in those
simulations in which both white  dwarfs are disrupted as a consequence
of a direct collision of two carbon-oxygen white dwarfs (runs 3 and 7,
respectively) the  masses of  $^{56}$Ni produced during  the extensive
nuclear burning  resulting from  the interaction are  of the  order of
$M_{\rm  Ni}\simeq 0.7 \,  M_{\sun}$.  Accordingly,  these simulations
result  in  Type Ia  supernovae  with  typical luminosities  ($L\simeq
1.6\times 10^{43}$~erg~s$^{-1}$).  In those runs in which the lightest
white dwarf is made of helium  and the direct collision results in the
disruption of  both stars (simulations 38,  40, and 44)  the amount of
synthesized  nickel  is  much  smaller,  typically  of  the  order  of
$10^{-3}\,  M_{\sun}$, as are  the corresponding  luminosities ($L\sim
10^{40}$~erg~s$^{-1}$),  and  would  not  be  classified  as  Type  Ia
supernovae. When  only one carbon-oxygen white dwarf  is disrupted and
ejected  (runs 4, and  8) the  masses of  nickel synthesized  are also
considerably  smaller ($M_{\rm  Ni}\simeq 0.06  \, M_{\sun}$)  and the
luminosities of these  events are, hence, smaller as  well ($L\sim 1.3
\times 10^{42}$~erg~s$^{-1}$).  Consequently, they would be classified
as sub-luminous supernovae.  As it  occurs in those cases in which the
lightest white dwarf  is made of helium and  both stars are disrupted,
in the simulations  in which only the helium  white dwarf is destroyed
and ejected (runs 37, 41, 42, 49, 53, and 54) the masses of nickel are
very small as well.  Finally, as  mentioned, in the rest of the cases,
namely those simulations  in which the interaction does  not result in
the  disruption  of  at  least  one  star,  the  masses  of  $^{56}$Ni
synthesized are negligible.

\subsection{Comparison with previous works}

\cite{RaSc10}   performed  simulations   of  head-on   and  off-center
collisions  with   initial  parameters   apparently  similar   to  the
simulations presented in this paper,  and found that $0.53\, M_{\sun}$
of nickel is produced in a $0.64\, M_{\sun}+0.81\, M_{\sun}$ collision
with impact parameter $b\simeq 1.0\, R_{\rm WD}$, whereas a negligible
amount  of nickel  is produced  in  a simulation  with $b\simeq  2.0\,
R_{\rm  WD}$,  which  results  in a  bound  remnant.   Our  equivalent
simulations  are runs  1, 5  and 9,  all of  them involving  two white
dwarfs of masses $0.6\,  M_{\sun}$ and $0.8\, M_{\sun}$, respectively.
All these simulations result in bound remnants and minuscle amounts of
nickel are produced.  The simulations  that most closely resemble each
other are  our run 5,  for which the  distance between the  centers of
mass of the two stars just  before the less massive white dwarf starts
transferring  mass to  the more  massive one  is $\simeq  1.9\, R_{\rm
WD}$, and  simulation 2  of \cite{RaSc10}  with $b\simeq  2.0\, R_{\rm
WD}$.  In both  simulations the outcome of the interaction  is a bound
remnant and  the amount of  nickel produced during the  interaction is
negligible.  Nevertheless,  to better compare with  the simulations of
\cite{RaSc10} we ran a set of additional simulations.  For this set of
runs we  adopt as a  fiducial model a  head-on collision of  two white
dwarfs of masses  $0.6\, M_{\sun}$.  We emphasize that  except for the
masses of  the colliding white dwarfs,  which in our case  are $0.60\,
M_{\sun}$ instead  of $0.64\,  M_{\sun}$, this fiducial  simulation is
identical to their run 1 with  $b=0$.  For this simulation we obtain a
mass  of  nickel  of  $0.22\, M_{\sun}$,  while  they  obtain  $0.51\,
M_{\sun}$.  However, we note that  the amount of $^{56}$Ni synthesized
depends very sensitively on several  factors.  Amongst them we mention
the  masses of  the interacting  white  dwarfs, how  the evolution  of
temperature is computed, the  resolution employed in the calculations,
and the adopted prescription for  the artificial viscosity. We discuss
them one by one.

To start with,  we note that the masses of  the colliding white dwarfs
are slightly different in both cases and, as mentioned previously, the
peak temperature reached during the interaction depends sensitively on
the  masses of  the  interacting white  dwarfs, as  does  the mass  of
synthesized  nickel.  The  second important  factor to  be taken  into
account is  that the peak  temperature reached during  the interaction
obviously depends on how the evolution of the temperature is computed.
We note  that for a  degenerate electron gas the  temperature obtained
from  the  energy  equation  ---  see  Sect.~\ref{input}  ---  may  be
incorrect by a  sizable percentage.  This is the reason  why for those
regions we adopt  a different formulation and we  follow the evolution
of  the temperature  using Eq.~(\ref{temp}),  which we  judge is  more
appropriate under these conditions.  Moreover, \cite{Dan12} have shown
that the peak  temperature and the averaged temperature  may differ by
up to a factor of $\sim 2$ --- see, for instance, their Fig.~4.  Given
the  extreme  sensitivity  of  the   nuclear  reaction  rates  to  the
temperature this,  quite naturally, translates in  large variations of
the  mass  of  nickel  synthesized.  For  instance,  in  our  fiducial
simulation we obtain a peak temperature $T_{\rm peak}\simeq 8.21\times
10^9$~K,  while \cite{RoKa09}  obtain  $T_{\rm peak}\simeq  8.90\times
10^9$,  which is  similar to  ours. However,  the masses  of $^{56}$Ni
synthesized in  both simulations differ considerably.   In particular,
we  obtain $0.22  \, M_{\sun}$,  whereas \cite{RoKa09}  obtain $0.32\,
M_{\sun}$.  Also the number of SPH particles plays a significant role.
To quantify this we ran an additional simulation in which we decreased
the number of SPH particles to $4\times  10^4$, that is by a factor of
5, and  we obtained that the  mass of nickel synthesized  in this case
was $0.094\,  M_{\sun}$.  Interestingly, \cite{RoKa09} find  that when
$2\times  10^6$  SPH  particles  are   employed  the  mass  of  nickel
synthesized  in the  explosion is  $0.32\, M_{\sun}$,  which is  quite
similar to  that found in  our fiducial simulation,  and \cite{RaSc10}
estimate that when  $5\times 10^4$ particles are used  the nickel mass
should be $\sim 0.3\, M_{\sun}$  which agrees relatively well with the
value  found in  our fiducial simulation.   Moreover, \cite{RoKa09}  find that
when  the  Eulerian  hydrodynamical  code FLASH  is  employed  in  the
calculations  the  mass of  nickel  is  considerably smaller,  $0.16\,
M_{\sun}$.  Thus, our mass of nickel  is bracketed by the values found
by \cite{RoKa09}.   The adopted treatment of  the artificial viscosity
also plays a non-negligible role. To illustrate this point we also ran
a series of low-resolution  simulations with $4\times 10^4$ particles,
in which the parameter $\alpha$  in Eq.~(\ref{av}) was varied from 0.5
to 1.0  and 1.5.  The  resulting nickel  masses are 0.017,  0.094, and
$0.076\, M_{\sun}$,  respectively.  Moreover, when the  Balsara switch
is not employed in the  calculations the mass of $^{56}$Ni synthesized
in  the  simulation with  $\alpha=0.5$  is  $M_{\rm Ni}\sim  \,  0.012
M_{\sun}$.   Thus,  depending  on  the  adopted  prescription  of  the
artificial viscosity the mass of nickel can  vary by up to a factor of
$\sim 5$.  In summary, we conclude that the mass of nickel synthesized
depends sensitively  on the details  of the numerical codes,  and that
discrepancies of  the order of  a factor of up  to $\sim 3$  can quite
naturally  arise   as  a   consequence  of  the   different  practical
implementations of the SPH formalism.

The natural question is now why we obtain such small amounts of nickel
in  some of  our simulations  when  compared with  the simulations  of
\cite{RaSc10}? The answer  to this question lays on the  choice of the
initial conditions.   We recall  that in  our simulations  the initial
orbits  correspond  to  a  post-capture  scenario  and  have  negative
energies, whereas in the simulations of \cite{RaSc10} the energies are
positive in most cases. Hence, our orbits are always elliptical, while
theirs  are in  most cases  either  parabolic or  hyperbolic. This  is
indeed at  the origin  of the  discrepancies in  the masses  of nickel
found  in both  sets  of  simulations.  To  illustrate  this point  we
compare our direct collision of two $0.8\, M_{\sun}$ white dwarfs (our
run 6)  with simulation 4  with $b=0$  of \cite{RaSc10}, in  which two
$0.81\, M_{\sun}$  white dwarfs collide  head-on.  We first  note that
the  impact  parameter  quoted  by \cite{RaSc10}  cannot  be  directly
compared  with that  given by  Eq.~(\ref{beta}) because  their initial
orbits  are open,  while ours  are elliptical.   In our  simulation we
obtain  a  bound  remnant,   while  \cite{RaSc10}  obtain  a  powerful
detonation  resulting in  the  total disruption  of  the system.   The
respective  masses   of  $^{56}$Ni   synthesized  are  in   this  case
$3.67\times  10^{-3}$  and  $0.84\,  M_{\sun}$.   However,  for  these
specific  simulations the  relative velocities  between both  stars at
contact are  very different.   In particular, the  relative velocities
are   $v_{\rm  rel}\simeq   4.5\times   10^{3}$~km/s  and   $5.9\times
10^{3}$~km/s,  respectively.   However,  while in  the  simulation  of
\cite{RaSc10}  this velocity  is  along the  line  connecting the  two
centers of  mass of the  white dwarfs, in  our case the  velocities of
each star form angles of $\sim\pm 33.6^\circ$ with the line connecting
the two centers of mass, thus resulting in a considerably less violent
collision, thus in a weaker shock,  and consequently in a much smaller
peak temperature and in a very small mass of synthesized nickel.

\section{Conclusions}
\label{concl}

In this paper we have studied  how the interactions of white dwarfs in
dense  stellar systems  depend on  the initial  conditions and  on the
masses  of the  intervening stars.   Our simulations  extend  those of
\cite{PaEn10}, in which the interactions of two white dwarfs of masses
$0.6$  and $0.8\,  M_{\sun}$  with different  initial conditions  were
studied, and encompass the most  plausible range of white dwarf masses
and  internal chemical compositions.   In total  we have  simulated 71
dynamical  interactions, of  which 36  correspond to  runs in  which a
regular  $0.8\, M_{\sun}$ interacts  with another  carbon-oxygen white
dwarf (of masses $0.6\, M_{\sun}$ and $1.0\, M_{\sun}$, respectively),
and a  $1.2\, M_{\sun}$ oxygen-neon white  dwarf.  In the  rest of the
simulations the  dynamical interactions  of a $0.4\,  M_{\sun}$ helium
white  dwarf with  either another  helium white  dwarf of  mass $0.2\,
M_{\sun}$,  or a  $0.8\,  M_{\sun}$ carbon-oxygen  white  dwarf, or  a
$1.2\, M_{\sun}$  oxygen-neon white dwarf were  explored.  Our initial
conditions have been  chosen to ensure that a  close encounter leading
to the formation of an eccentric binary or a collision always happens.

We have  found that the  outcome of the  interactions can be  a direct
collision, a lateral collision or the formation of an eccentric binary
system.  In direct  collisions there is only one  violent and dramatic
mass transfer episode in which the less massive white dwarf is tidally
disrupted by the more massive one  on a dynamical time scale, while in
a lateral  collision although  the less massive  star is  disrupted as
well, it takes  several orbits around the more  massive white dwarf to
be totally destroyed. Thus, in this case the entire disruption process
occurs in a more gentle  way.  Moreover, we have demonstrated that the
outcome of the interaction can  be predicted using very basic physical
principles.  In particular, we have  found that for a given simulation
tidal forces modify the  initial trajectories of the interacting stars
in such  a way  that the distance  at closest approach  determines the
final  outcome of  the dynamical  interaction.  Specifically,  we have
found  that if the  distance at  closest approach  is small  enough to
allow a deep contact between  both white dwarfs, a direct collision is
the natural outcome of the interaction.  For this to occur the overlap
between both stars at minimum distance must be of the order of 35\% if
two typical carbon-oxygen white  dwarfs are considered.  Else, we have
demonstrated  as well that  lateral collisions  occur when  at minimum
distance the  radius of the less  massive white dwarf  is within $\sim
0.95$ of the Roche lobe radius of the interacting system.

We  have  also  characterized  for  which initial  conditions  of  the
dynamically interacting system the material flowing from the disrupted
less massive  white dwarf and accreted  onto the more  massive star is
compressed  to  such an  extent  that  reaches  the conditions  for  a
detonation  to  develop. Moreover,  we  have  also  studied for  which
initial conditions the  explosion is powerful enough to  result in the
disruption of  one or  both of the  interacting white dwarfs,  and for
which ones degeneracy is lifted and the result of the interaction is a
central,  more massive  and very  hot  object surrounded  by a  debris
region  orbiting  around  it.    Our  results  indicate  that  if  the
intervening  stars are  regular  carbon-oxygen white  dwarfs (or  even
oxygen-neon  ones) detonations occur  when the  two components  of the
system are separated  by less than $\sim 0.015\,  R_{\sun}$ at minimum
distance, and  that one or both  components of the  system are totally
disrupted and ejected  to the surrounding medium if  the total mass of
the system  is rather large,  preferentially $\ga 1.4\,  M_{\sun}$ and
the  initial  periastron  distance   is  smaller  than  $\sim  0.005\,
R_{\sun}$ --- see Fig.~\ref{plane}.  However, if the less massive star
is a helium  white dwarf the resulting detonations  always result in a
catastrophic output  for separations at minimum distance  $\la 0.02 \,
R_{\sun}$. Two of our simulations result in a super-Chandrasekhar Type
Ia  supernova  outburst, corresponding  to  direct  collisions of  two
rather massive  carbon-oxygen white dwarfs of  masses $1.0\, M_{\sun}$
and $0.8\,  M_{\sun}$.  There are as  well a few  simulations in which
only  one carbon-oxygen white  dwarf is  disrupted and  ejected. These
simulations also result in powerful explosions, but their luminosities
are considerably  smaller and, thus,  would probably be  classified as
sub-luminous  supernovae.    Finally,  other  simulations   result  in
powerful outbursts, and  lead as well to the  disruption of the entire
system, but  involve white dwarfs  with helium cores.  In  these cases
the  mass  of  $^{56}$Ni  is  very small,  as  are  the  corresponding
luminosities.   Nevertheless, some  of our  simulations  produce bound
remnants  which  are  close   to  the  Chandrasekhar  limit,  and  the
subsequent evolution  of these systems may eventually  produce Type Ia
supernovae,  as  the  viscous  evolution  unbinds only  a  very  small
fraction of the material of the debris region \citep{Schwab12}.

For those  interactions resulting in  a central, massive and  very hot
remnant  we have  also  studied the  influence  of the  masses of  the
interacting  white  dwarfs  and  of  the  initial  conditions  on  the
properties of the final remnants.   In particular, we have studied the
morphology  of the  resulting remnant,  the peak  temperatures reached
during the most  violent phase of the interaction,  and the associated
nucleosynthesis.  In  all these simulations a central  hot white dwarf
surrounded by a debris region  is formed, whereas a variable amount of
mass is ejected  from the system. The morphology  of the debris region
depends  mostly on  the kind  of collision  the system  undergoes.  In
particular,  for lateral  collisions  the debris  region  is a  heavy,
rotationally-supported  keplerian  disk, whilst  for  direct ones  the
debris regions consists of  a spheroid. The peak temperatures attained
during the most violent phase of the accreting episode are rather high
in all  cases, typically  of the  order of $10^9$~K,  and can  be even
larger  for  direct  collisons,  for which  temperatures  larger  than
$10^{10}$~K are easily reached,  than for lateral interactions.  This,
in  turn, drives  extensive  nucleosynthetic activity  in the  shocked
regions.  Hence, the debris region and the material ejected during the
dynamical interaction are substantially enriched in heavy elements.

In summary, we have computed  a comprehensive set of simulations aimed
to   provide  a   consistent  framework   to  analyze   the  dynamical
interactions  of  white  dwarfs   in  dense  stellar  systems.   These
interactions  are  of interest  because  the  collision is  likely  to
detonate the white dwarfs, and result in a type Ia supernova outburst.
Actually, in our simulations the  detonation conditions are reached in
a significant number of interactions,  and the masses of the exploding
systems show  some dispersion. As a  matter of fact, we  find that for
some  simulations the  detonation  occurs in  interacting systems  for
which the involved  mass is larger than  Chandrasekhar's mass, whereas
in  some  other  the  resulting explosion  is  sub-Chandrasekhar.   An
important fact that needs to be taken into account is that it has been
recently  shown  that such  interactions  might  be more  common  than
previously  thought \citep{KD12},  and could  even dominate  the event
rate.   Thus,   there  is  a   renewed  interest  in   studying  these
interactions, for  which there was  a lack of  extensive calculations.
Precisely,  our results  fill  this gap,  and pave  the  road to  more
extensive  calculations in  which the  enhancement in  the event  rate
proposed   by    these   authors    could   be   analyzed    in   more
detail. Particularly, it is worth mentioning that this increase in the
event rate depends on the fraction  of white dwarf collisions that are
expected to lead  to Type Ia-like events. Given  that our calculations
provide  the maximum  separation at  pericenter for  various pairs  of
white  dwarfs  to  produce   detonations,  future  calculations  could
approach this problem on a solid basis. Finally, another open question
remains to be  answered yet, namely how exactly  the optical signature
of the  explosions resulting  from these dynamical  interactions would
look like, even if they do not result in Type Ia supernovae. Given the
wide range of  nickel masses produced during  the interactions studied
here,  they  may  explain   both  some  underluminous  transients  and
relatively bright outbursts.  This is,  nevertheless, out of the scope
of this paper, and should be studied in future works.

\section*{Acknowledgments}
This  work was  partially  supported by  MCINN grants  AYA2011--23102,
AYA2010--15685 and AYA2011--24780, by the AGAUR, by the European Union
FEDER funds, and by  the ESF EUROGENESIS project (grants EUI2009-04167
and EUI2009-04170).

\bibliographystyle{mn2e}
\bibliography{grid_low}

\end{document}